\documentclass{emulateapj}

\usepackage{amssymb,amsmath,graphicx} 

\newcommand\beq{\begin{equation}}
\newcommand\eeq{\end{equation}}
\newcommand\beqar{\begin{eqnarray}}
\newcommand\eeqar{\end{eqnarray}}

\newcommand{\tni}{\tau_{\rm ni}}

\newcommand{\cref}{C_{\rm ref}}

\begin{document}

\title{Non-equilibrium Chemistry of Dynamically Evolving Prestellar Cores:\\ I.  Basic Magnetic and Non-Magnetic Models 
and Parameter Studies}

\author{Konstantinos Tassis\altaffilmark{1,2}, 
Karen Willacy\altaffilmark{1},
Harold W. Yorke\altaffilmark{1},
\& Neal J. Turner\altaffilmark{1}}

\altaffiltext{1}{Jet Propulsion Laboratory, California Institute of Technology, Pasadena, CA 91109, USA}
\altaffiltext{2}{current address: Max-Planck Institute for Radio  Astronomy, 53121 Bonn, Germany}

\begin{abstract}

We combine dynamical and non-equilibrium chemical modeling of evolving prestellar molecular cloud cores
and investigate the evolution of molecular abundances in the contracting core. We model both magnetic cores, with 
varying degrees of initial magnetic support, and non-magnetic cores, with varying collapse delay
times. We explore, through a parameter study, the competing effects of various model parameters in the 
evolving molecular abundances, including the elemental C/O ratio, the temperature, and the cosmic-ray ionization rate. 
We find that different models show their largest quantitative differences at the center of the core, 
whereas the outer layers, which evolve slower, have abundances which are severely degenerate among 
different dynamical models. There is a large range of possible abundance values for different models 
at a fixed evolutionary stage (central density), which demonstrates the large potential of chemical differentiation in prestellar cores. However, degeneracies among different models, compounded with uncertainties induced by other model parameters, make it  difficult to discriminate among dynamical models. To address these difficulties, we identify  abundance 
ratios between particular molecules, the measurement of which would have maximal potential for 
discrimination among the different models examined here. In particular, we find that the ratios between NH$_3$ and CO; NH$_2$ and CO; NH$_3$ and HCO$^+$ are sensitive to the evolutionary timescale, and that the ratio between HCN and OH
is sensitive to the C/O ratio. Finally, we demonstrate that measurements of the central deviation (central depletion or enhancement) of abundances of certain molecules are good indicators of the dynamics of the core. 
\end{abstract}


\section{Introduction}

The earliest phases of star formation have been the subject of intense observational and 
theoretical investigation over the past four decades. Nevertheless, together with the very 
late stages of star formation (formation of planetary systems), the very early processes of 
molecular cloud fragmentation into high-density ``cores'' and the subsequent early contraction of these cores
before protostars form in their centers remain the two least constrained and most debated aspects 
of star formation theory (McKee and Ostriker 2007). The reason for this lack of strong  constraints 
is that it is difficult to extract physical conditions in star forming regions from  observable 
quantities, such as dust column or molecular line maps; each observable is typically affected  by 
multiple factors which are nontrivial to deconvolve. As a result, theoretical models of cloud  
fragmentation and core evolution span a large range of possibilities. 

A diverse array of mutually interacting microphysical processes influence the evolution of molecular 
clouds to differing extents:  gravity, turbulence, magnetic fields, UV and cosmic ray (CR) ionization, and a 
complex network of chemical reactions. Without strong observational constraints on the relative 
importance of these processes, competing star formation theories have arisen based on different 
assumptions of the physics driving the formation of prestellar cores and their collapse to form protostars.

The path to the appearance of a protostar depends sensitively on the initial conditions and the 
relative importance of the different physical processes driving the formation and contraction of 
prestellar molecular cloud cores.The most important distinction among different calculations of 
prestellar collapse pertains to the nature of any forces opposing gravity during the prestellar 
collapse. In this context, we can identify two broad classes of models: non-magnetic and magnetic.

Non-magnetic collapsing models include early numerical and semi-analytic solutions of the Euler equations 
of hydrodynamic collapse (e.g., Larson 1969; Penston 1969; Yorke \& Kr\"ugel 1977; Shu 1977)
\footnote{Note however that Shu's singular isothermal sphere models refer to protostellar cores in which a 
central object exists, since the initial conditions for the inside-out collapse are singular at the center. 
Thus, this class of models cannot be used to study the early prestellar phase of the core evolution.}, 
as well as numerical studies of the gravo-turbulent formation and evolution of bound condensations formed 
by converging flows in turbulent model clouds (see Mac Low \& Klessen 2004 and Scalo \& Elmegreen 2004 for recent 
reviews). The evolution of cores formed by turbulence-driven fragmentation is compatible with models of 
pure hydrodynamical collapse  (e.g., Gong \& Ostriker 2009). Turbulence-initiated core formation models 
can be generally separated into models of rapid global collapse, where the entire molecular cloud is 
viewed as an evanescent structure (e.g., Hartmann et al.\ 2001; Heitsch \& Hartmann 2008), and models 
with some initial decaying turbulent support that extends the lifetime of clouds 
(e.g., Krumholz et al.\ 2006; Tan et al.\ 2006). 

Magnetic models include magnetohydrodynamical calculations of the magnetically mediated 
contraction of prestellar cores, which is typically slower than the pure hydrodynamical collapse 
(with lower infall velocities and longer evolution timescales, e.g.\ Fiedler \& Mouschovias 1993; 
Li \& Shu 1996; Tassis \& Mouschovias 2007), and result in a characteristic 
hourglass-like configuration of the magnetic field threading the core 
(see Mouschovias \& Ciolek 1999 for a review). In magnetic models, the lifetime of the parent molecular 
cloud is controlled by the initial value of the mass-to-magnetic-flux ratio. If this ratio is smaller 
than a critical value (the cloud is {\em magnetically subcritical}), then the cloud as a whole is 
long-lived as the magnetic field can support it against gravity. In this case, 
{\em magnetically supercritical} cores form through the process of ambipolar diffusion and collapse to 
form protostars, while the cloud envelope remains supported by the field. If the mass to magnetic flux ratio 
of the parent cloud is larger than the critical value, then the cloud as a whole is unstable and 
short-lived, while the formation and collapse of cores also proceeds in shorter timescales.

The coupling between chemistry and dynamics is an essential step in connecting the dynamics of prestellar 
cores to observations, and has long been recognized as critical in extracting information from molecular 
line observations. The problem has been pursued for the past decade in a series of ground-breaking studies.
Bergin \& Langer (1997) identified, through studies of static media, ``early'' and ``late'' 
molecules (molecules which peak in abundance either in the initial or in the later stages of the evolution 
of a molecular cloud fragment). Aikawa et al. (2001) compared observations of L1544 with coupled 
chemical/dynamical models. For the dynamics, they used the analytical similarity solution of Larson (1969) 
and Penston (1969), as well as a ``slowed down'' version of it (using an overall slow-down factor) to 
approximate the slower magnetic collapse, and found that their fast models are better fits to L1544. 
Li et al. (2002) and Shematovich (2003) also modeled L1544, using a non-magnetic model and the Li (1999) 
magnetic model for the dynamics, which treats the magnetic field in an approximate way including only 
magnetic pressure forces and assuming a spherical geometry. They also included grain-surface reactions 
in their calculations, and they found the magnetic model to be a better fit to the observations. 
They showed that the magnetic collapse is accelerating, and therefore a uniform slow-down of the 
non-magnetic collapse is not a good approximation, because in magnetic collapse the amount of time spent 
at different densities changes with respect to the non-magnetic case. Aikawa et al. (2003) updated the 
Aikawa et al. (2001) calculation to include grain-surface reactions, and they found that moderate 
retardation factors (factor of 3) are not in as strong disagreement with the data as in 
Aikawa et al. (2001), however they still found that the faster models are preferred. 
Li et al. (2002) however noted that their preferred Larson-Penston solution features very large infall 
velocities that cannot be reconciled with observations of L1544.  Lee et al. (2003) modeled the chemistry 
of various prestellar cores to fit Bonnor-Ebert and Plummer-like radial density profiles using dust 
continuum and line emission. Lee et al. (2004) looked at the chemical evolution of prestellar cores 
aiming to produce line profiles for prestellar and protostellar cores. For modeling the dynamical 
evolution of prestellar cores they used a sequence of Bonnor-Ebert profiles, however the 
assignment of times and velocities to each profile was not treated self-consistently.  For the protostellar 
cores they used the inside-out collapse model of Shu (1977).  Flower et
al. (2005), using a homologous collapse treatment of prestellar cores,
emphasized the effect of grain growth on the chemistry. Aikawa et al. (2005) revisited the prestellar 
core problem, but instead of using an analytic solution for the dynamics, they used  post-processing of a 
numerical simulation of the collapse of Bonnor-Ebert spheres  (one very close to equilibrium and one very 
unstable), and found significant differences in the system chemistry depending on the initial conditions. 
This result was confirmed by Keto \& Caselli (2008) who coupled one-dimensional dynamical simulations with a 
simple CO chemistry network and radiative transfer. In a follow-up study, Keto \& Caselli (2010), starting from a  Bonnor-Ebert sphere of central density $2\times 10^{4} \rm {\, cm^{-3}}$ in unstable equilibrium, evolving quasi-statically, found that at a density of $2\times 10^{7} \rm {\, cm^{-3}}$ the abundance of CO is much lower than observations in L1544, and argued for a higher CO desorption rate from grains. 

On certain issues a consensus is forming among these studies 
(see, e.g., Bergin \& Tafalla 2007 for a recent review). One such issue is the sensitivity of the chemical 
state of the contracting core to the duration of the collapse. Characteristic examples are the 
carbon-bearing molecules, which tend to become depleted on the ice mantles of dust grains and are therefore 
most prominent in the early stages of evolution of a core, in contrast to nitrogen-bearing molecules which 
are not depleted and, as a result, their abundance increases in later stages.

However, core age is not the only contributing factor in determining molecular abundances.  
Observations have established the existence of chemical differentiation in prestellar cores of otherwise 
similar properties, which however show strong variations in the abundances of a number of chemical species.
Characteristic examples of  chemical differentiation can be seen in observations of prestellar cores 
L1544, L1498, L1517B  which show CO depletion (Tafalla et al. 2002, 2004), while L1521E 
(a core of similar central density with L1517B) shows no CO depletion (Tafalla \& Santiago 2004). This 
differentiation may be a result of the dynamics of the collapse as suggested by, e.g.,  
Shematovich et al. (2003), and thus reflect the initial conditions in the parent star forming region.

In this paper, we self-consistently follow the chemistry and dynamics of contracting prestellar cores. 
We do so by adding to the system of equations of Eulerian hydrodynamics, or MHD (as appropriate), 
one continuity equation with sources and sinks given by the chemical reactions network 
{\em for each species} followed. In this way, the dynamics and non-equilibrium chemistry evolve 
simultaneously. This is particularly important in the case of MHD models, as abundances 
of charged species have a feedback effect on the dynamics, since the magnetic field couples only to 
charged molecules and grains, which in turn determine the evolution of the magnetic field and thus 
the dynamics of the contraction. In a companion publication (Tassis et al.~2011, in preparation, hereafter Paper II), we discuss in more detail the interplay between chemistry and dynamics in magnetic models.

Our paper is organized as follows. In \S \ref{dynamics} we discuss our dynamical models 
(hydrodynamic and MHD) of core evolution. The chemical network we implement is discussed 
in \S \ref{chemistry}. Details on the parameter study we have conducted are given in \S \ref{parstu}. 
We present our results in \S \ref{res} and we discuss our findings and conclusions in \S \ref{thedisc}.

\section{Dynamical Models of Molecular Cloud Cores}\label{dynamics}

\subsection{Hydrodynamic Core Models}
We model the contraction of a non-magnetic, self-gravitating, isothermal prestellar core in spherical 
symmetry. The core is assumed to be embedded within a uniform-density isothermal cloud. We solve the  
equations of hydrodynamics in spherical symmetry: 
\begin{subequations}
\beq
\frac{\partial \rho}{\partial t} + \frac{1}{r^2} 
\frac{\partial (r^2 \rho v_r)}{\partial r}=0 ,\label{HDcont}
\eeq
\beq
\frac{\partial \rho_j}{\partial t} + \frac{1}{r^2} 
\frac{\partial (r^2 \rho_j v_r)}{\partial r}=S_j ,\label{HDcontChem}
\eeq
\beq
\frac{\partial (\rho v_r)}{\partial t} +
\frac{1}{r^2} \frac{(\partial r^2 \rho v_r^2)}{\partial r} = 
-C_{\rm s}^2 \frac{\partial \rho}{\partial r} + \rho g_r , \label{HDmom}
\eeq
\beq
g_r=-\frac{4 \pi G}{r^2} \int_0^r \! \rho r'^2 \, dr',
\eeq
\end{subequations}
where $v_r$ is the radial velocity, $\rho$ is the total density of the gas, and 
$\rho_j$ is the density of each individual chemical species followed (listed in Table \ref{table1}).
We solve a density continuity equation for each of the chemical species in the model 
(equation \ref{HDcontChem}). In these equations, $S_j$ are source/sink terms describing
mass interchange between the species due to chemical reactions. 

The initial state is one of uniform density equal to $10^3 {\rm
    \, cm^{-3}}$ and zero velocity. The radius of the core is $0.4$
  pc, making it significantly thermally supercritical. At the outer boundary, we
  require constant pressure and zero velocity. This way, the core
  experiences no influx of mass.

\subsection{MagnetoHydroDynamic Core Models}

We follow Basu \& Mouschovias (1994) in modeling the formation and contraction of prestellar cores in
self-gravitating, magnetic, rotating, isothermal model clouds in the presence of ambipolar diffusion
and magnetic braking. In the presence of an ordered magnetic field the cloud relaxes rapidly along field 
lines until thermal-pressure forces balance gravity, before any significant contraction perpendicular 
to the field lines takes place (Fiedler \& Mouschovias 1993; Desch \& Mouschovias 2001; 
Kunz \& Mouschovias 2010). Force balance along magnetic field lines is maintained even when the contraction 
becomes dynamic in the perpendicular direction. Consequently, the thin-disk approximation is adopted for 
the model cloud cores (Ciolek \& Mouschovias 1993). The model cloud is axisymmetric with the rotation axis 
coinciding with the axis of symmetry, which is taken to be the z-axis of a cylindrical polar coordinate 
system ($r,\phi,z$). The upper and lower surfaces of the disk are located at $z=\pm Z(r,t)$. The cloud is 
embedded in an ``external'' medium of constant thermal pressure $P_{\rm ext}$, and density $\rho_{\rm ext}$, 
and is threaded by a magnetic field which, far from the cloud, becomes uniform and equal to a ``reference'' 
value $B_{\rm ref}$.

The thin disk evolution equations, obtained by integrating the two-fluid (plasma and neutrals) MHD equations 
over $z$, are
\begin{subequations}
\beq\label{thin1}
\frac{\partial \sigma_{\rm n}}{\partial t} + \frac{1}{r}\frac{\partial (r 
\sigma_{\rm n} v_{{\rm n},r})}{\partial r}  =  0, 
\eeq  
\beq\label{thin2}
\begin{split}
\frac{\partial (\sigma_{\rm n} v_{{\rm n},r})}{\partial t} + \frac{1}{r}\frac{\partial( r 
\sigma_{\rm n} v_{{\rm n},r}^2)}{\partial r} & =  -C_{\rm eff}^2 \frac{\partial
\sigma_{\rm n}}{\partial r} + \sigma_{\rm n} g_r \\ 
&+\frac{L_{\rm n}^2}{\sigma_{\rm n} r^3}+ F_{{\rm M},r} ,
\end{split}
\eeq
\beq\label{thin3}
\frac{\partial L_{\rm n}}{\partial t} + \frac{1}{r}\frac{\partial( r 
L_{\rm n} v_{{\rm n},r})}{\partial r}  =  \frac{1}{2\pi} r B_{z,{\rm eq}} B_{\phi,Z},
\eeq
\beq\label{thin4}
\frac{\partial B_{z,{\rm eq}}}{\partial t} + \frac{1}{r}\frac{\partial (r 
B_{z,{\rm eq}} v_{{\rm i},r})}{\partial r}  =  0, 
\eeq
\beq\label{thin5}
\rho_{\rm n}(r,t)C^2 = P_{\rm ext} + \frac{\pi}{2}G \sigma_{\rm n}^2(r,t),
\eeq
\beq\label{thin6}
\sigma_{\rm n}(r,t) = \int_{-Z(r,t)}^{+Z(r,t)} \rho_{\rm n} dz  =  2
\rho_{\rm n}(r,t)Z(r,t), 
\eeq
\beq\label{thin7}
C_{\rm eff}^2 = \frac{\pi}{2}G \sigma_{\rm n}^2 \frac{[3P_{\rm ext} + (\pi/2)G
  \sigma_{\rm n}^2]}{[P_{\rm ext} + (\pi/2)G\sigma_{\rm n}^2]^2} C^2,
\eeq
\beq\label{thin8}
\begin{split} 
F_{{\rm M},r} &= \frac{1}{2\pi} \left\{ B_{z,{\rm eq}}\left( B_{r,Z}-Z\frac{\partial
  B_{z,{\rm eq}}}{\partial r}\right) + \frac{1}{2} \frac{\partial Z}{\partial r} \right .\\
&\times \left[ B_{r,Z}^2 + 2B_{z,{\rm eq}}\left(B_{r,Z}\frac{\partial Z}{\partial r}\right)  \right .\\
&\left. \left. +B_{\phi,Z}^2 +\left(B_{r,Z}\frac{\partial Z}{\partial r}\right)^2 \right]  \right\}, 
\end{split}
\eeq 
\beq\label{thin9}
B_{\phi,Z}(r) = - 2 \frac{(4\pi \rho_{\rm ext})^{1/2}}{B_{\rm ref}} \frac{\Phi}{r} (\Omega-\Omega_{\rm b}),
\eeq
\beq\label{thin10}
B_{r,Z}(r) = - \int_{0}^{\infty}dr' r' [B_{z,{\rm eq}}(r')-B_{\rm ref}]\mathcal{M}(r,r'),
\eeq
\beq\label{thin11}
g_{\rm r}(r) = 2\pi G \int_{0}^{\infty}dr' r' \sigma_{\rm n}(r')\mathcal{M}(r,r'),
\eeq
\beq\label{thin12}
\begin{split}
\mathcal{M}(r,r') =& \frac{d}{dr}\left[ \int_{0}^{\infty} dk J_0(kr)J_0(kr')
\right]\\
=&\frac{2}{\pi}\frac{d}{dr}\left[ \frac{1}{r_>}K\left( \frac{r_<}{r_>}\right) \right],
\end{split}
\eeq
\beq\label{thin13}
\frac{\partial \sigma_{\rm j0}}{\partial t} + \frac{1}{r}\frac{\partial (r 
\sigma_{\rm j0} v_{{\rm n},r})}{\partial r}  =  S_{\rm j0}, 
\eeq
\beq\label{thin14}
\frac{\partial \sigma_{\rm j+}}{\partial t} + \frac{1}{r}\frac{\partial (r 
\sigma_{\rm j+} v_{{\rm i},r})}{\partial r}  =  S_{\rm j+}, 
\eeq    
\beq\label{thin15}
v_{{\rm i},r} = v_{{\rm n},r} + \frac{<\tni>}{\sigma_n}F_{{\rm M},r},
\eeq
\end{subequations}
%%%%%%%%%%%%%%%%%%%%%%%%%%%%%%%%%%%%%%%%%%%%%%%%%%%%%%%%%%%%%%%%%%%%%%%%%%%%%%%%%%%%%%%
\noindent
where $\sigma_{\rm n}$ and $\rho_{\rm n}$ are the column and volume densities of neutrals respectively,
$v_{{\rm i},r}$ and $v_{{\rm n},r}$ are the radial ion and neutral velocities respectively,
$L_{\rm n}=\sigma_{\rm n} v_{{\rm n},\phi} r$ is the angular momentum
per unit area, and 
$\Phi=\int^r_0 dr' r'B_{z,{\rm eq}}$ is the magnetic flux.  
Equation (\ref{thin1}) is the mass continuity equation of the neutrals. 
Equation (\ref{thin3}) gives the $\phi$ component of the neutral force equation, while the 
$r$-component is given by Equation (\ref{thin2}). Equation (\ref{thin5}) states that the thermal 
pressure at the equatorial plane in each magnetic flux tube balances the external and gravitational 
pressures. The introduction of the effective sound speed, $C_{\rm eff}$, accounts
for the contribution of the radial force exerted by the external pressure
$P_{\rm ext}$ on the upper and lower surfaces of the flaring disk, since the latter
are not horizontal. Furthermore, in the expression for the total radial
magnetic force (per unit area), $F_{{\rm M},r}$, the first two terms are the
magnetic tension force and the magnetic pressure force within the cloud, respectively.
The remaining terms are corrections due to the variation of the half-thickness of the 
cloud, $Z$, with $r$. Equation (\ref{thin4}) is Faraday's law in the two-fluid approximation with the magnetic flux frozen in the ion fluid.
The toroidal component of the magnetic field ($B_{\phi,Z}$) is determined assuming  
a steady state condition on the transport of angular momentum by torsional Alfv{\'e}n waves 
in the external medium and given by equation (\ref{thin9}).
The radial component of the magnetic field at the top surface of 
the thin disk ($B_{r,Z}$) is given by equation (\ref{thin10}), obtained by assuming a force-free 
external medium, whereby the half-thickness of the
disk ($Z$) is obtained from equation (\ref{thin6}) after $\rho_{\rm n}(r,t)$ is
calculated from equation (\ref{thin5}). The $r$-component of the gravitational field (\ref{thin11})
is calculated from Poisson's equation for a thin disk. $J_0$ is the 
zeroth-order Bessel function of the first kind, $K$ is the complete elliptic integral of 
the first kind, $r_< = {\rm min}[r,r']$, and $r_> = {\rm max}[r,r']$.    
Equation (\ref{thin13}) is the mass continuity equation for neutral gas phase 
chemical species advected with the neutral bulk velocity $v_{{\rm n},r}$. 
Equation (\ref{thin14}) is the mass continuity equation for the gas phase ions and all
the grain mantle species (since dust grains are overwhelmingly negatively charged in the
densities we are studying) advected with the velocity of the ions $v_{{\rm i},r}$. 
$S_{\rm j0}$ and $S_{\rm j+}$ represent the source/sink terms due to chemical reactions. 
Note that the ions are advected with a different velocity than the neutrals, so 
two equations are required in the magnetic model to replace Eq.~(\ref{HDcontChem}) of the non-magnetic model.

Equation (\ref{thin15}) gives the velocity of the ions and is obtained from the algebraic solution of
the force equation for the ion fluid where the acceleration terms have been neglected because of the small
inertia of these species. The mean (momentum exchange) collision time of a neutral particle with ions 
in equation (\ref{thin15}) is
\beq
<\tau_{\rm ni}>=1.56\frac{<\mu_{\rm i}>m_{\rm p}+2m_{\rm p}}{\rho_{\rm i} <\sigma w>_{\rm in}},
\eeq
where $m_{\rm p}$ is the proton mass, $\rho_{\rm i}=\sum_{\rm j+} \rho_{\rm j+}$ is the total ion density, 
the mean molecular weight of the ions is 
\[<\mu_{\rm i}>=\sum_{\rm j+} m_{\rm j+}n_{\rm j+}/\sum_{\rm j+} n_{\rm j+}\,,\]
$n_{\rm j+}$ is the number volume density for each one of the ions considered, and
$<\sigma w>_{\rm in}$ is the average collisional rate between ions and hydrogen molecules.
The value of $<\sigma w>_{\rm in}$ varies slightly with the mass of the ion (McDaniel \& Mason 1973); 
we use the canonical 
value $<\sigma w>_{\rm in}=1.69\times 10^{-9}$ ${\rm cm^3s^{-1}}$ (Ciolek \& Mouschovias 1993).

Each magnetic model is  initiated at a reference state as in Basu
  \& Mouschovias (1994). The reference state has a column density
  profile of the form
\begin{equation}
\sigma_{\rm n,ref}(r) = \frac{\sigma_{\rm c,ref}}
{\left[ 1+(r/l_{\rm ref})^2\right]^{3/2}}\,,
\end{equation}
which implies an almost uniform column density for $r\ll l_{\rm ref}$,
which is where the core forms. The high-$r$ tail has little effect on
the calculations and is there for numerical convenience, allowing the
calculation of the disk gravitational potential to converge. The
profile of the reference state angular velocity is obtained by using a
linear dependence of specific angular momentum on mass, corresponding
to the angular momentum distribution of a uniformly rotating disk, yielding
\begin{equation}
\Omega_{\rm ref}(r) = 2\Omega_{\rm c,ref}\left(\frac{l_{\rm
      ref}}{r}\right)^2
\left[1-\frac{1}{\sqrt{1+(r/l_{\rm ref})^2}}\right]\,.
\end{equation}
The reference state is  first allowed to relax to an equilibrium
configuration, while ambipolar diffusion and chemistry are turned
off.  It is from this equilibrium state that the subsequent evolution is 
initiated. For this reason, the initial central density to which each model relaxes is not always the same 
but it depends on the initial mass to flux ratio (see
Fig.~\ref{n_vs_t}). 

The free parameters in the magnetic models are the central density of
the initial reference state (which we take to be $10^3 {\rm \,
  cm^{-3}}$); the background angular velocity $\Omega_b = 10^{-15}
{\rm \, rad \, s^{-1}}$, approximately corresponding to the rate of
galactic rotation in the solar neighborhood;  the reference state angular velocity ($\Omega_{\rm ref}
= 0.48 {\rm \, km \, s^{-1} \, pc^{-1}}$); the external pressure
$P_{\rm ext} = 0.1 \pi G \sigma^2_{\rm c,ref}/2$ corresponding to
$10\%$ of the central ``gravitational'' pressure; the characteristic
size of the column density profile $l_{\rm ref} = 1 {\rm \, pc}$; the
external density $\rho_{\rm ext} = 10 {\rm cm^{-3}}$; and
the mass to magnetic flux ratio, which is varied (see discussion of
parameter study below). 

With the exception of the magnetic mass to flux ratio, the dynamical
evolution of the collapsing core is not sensitive to any of the free
parameters (Basu \& Mouschovias 1995).

 At the outer boundary of the disk we require continuity of thermal as
well as magnetic pressures (see Tassis \& Mouschovias 2005, \S 5). No
influx of mass is allowed at the outer boundary, so the total mass of
the cloud is conserved.

%%%%%%%%%%%%%%%%%%%%%%%%%% TABLE 1 %%%%%%%%%%%%%%%%%%%%%%%%%%%%%%%%%%%%%
\begin{table*}
\begin{center}
%{\large Table 1}\\
\caption{\label{table1} Chemical Species Considered}
\resizebox{\textwidth}{!}{
\begin{tabular}{l l l l l l l l l l l l l l l l}
\tableline\tableline
\multicolumn{16}{c}{gas phase species}\\%\\ \hline
\tableline
${\rm H^+}$   & H     & $\rm H_2^+$     & ${\rm H_2}$ & ${\rm H_3^+}$ & He   & ${\rm He^+}$   
& ${\rm C^+}$ & C    & CH   & ${\rm CH^+}$ & ${\rm CH_2^+}$  & ${\rm CH_2}$   
& N    & ${\rm N^+}$     & ${\rm CH_3}$    \\   
${\rm NH^+}$  & ${\rm CH_3^+}$ & NH & ${\rm NH_2^+}$  & O  & ${\rm CH_4}$ & ${\rm CH_4^+}$   
& ${\rm O^+}$ & ${\rm NH_2}$  & ${\rm CH_5^+}$ & OH & ${\rm OH^+}$  & ${\rm NH_3^+}$  
& ${\rm NH_3}$  & ${\rm H_2O}$  & ${\rm NH_4^+}$   \\
${\rm H_2O^+}$ & ${\rm H_3O^+}$ & ${\rm C_2}$ & ${\rm C_2^+}$ & ${\rm C_2H^+}$ & ${\rm C_2H}$ 
& ${\rm C_2H_2^+}$ & ${\rm C_2H_2}$ & CN & ${\rm CN^+}$ & ${\rm HCN^+}$ & ${\rm C_2H_3^+}$ & HCN   
& HNC  & ${\rm Si^+}$ & ${\rm C_2H_4^+}$  \\
${\rm H_2NC^+}$ & Si    & ${\rm N_2}$ & ${\rm CO^+}$ & ${\rm HCNH^+}$  & CO  & ${\rm N_2^+}$ & HCO
& ${\rm N_2H^+}$ & ${\rm HCO^+}$   & ${\rm H_2CO}$   & ${\rm H_2CO^+}$ & NO & ${\rm NO^+}$  
& ${\rm H_3CO^+}$  & ${\rm CH_3OH}$  \\
${\rm O_2}$   & ${\rm O_2^+}$   & ${\rm CH_3OH_2^+}$ & ${\rm C_3^+}$ & ${\rm C_3H^+}$ & ${\rm C_2N^+}$   
& ${\rm CNC^+}$ & ${\rm C_3H_3^+}$ & ${\rm CH_3CN}$ & ${\rm C_3H_2}$ & ${\rm CO_2}$ & ${\rm CO_2^+}$  
& ${\rm HCO_2^+}$ & ${\rm HC_3N}$ &       \\
\tableline
\multicolumn{16}{c}{grain mantle species}\\%\\ \hline
\tableline
H &  C   & CO   & ${\rm H_2CO}$   & Si    & ${\rm C_2}$    & ${\rm O_2}$ & CH & OH & NO & ${\rm CH_2}$   
& ${\rm H_2O}$ & ${\rm CO_2}$  & ${\rm CH_3}$  & ${\rm CH_4}$ & HNC     \\
${\rm C_2H_2}$ & ${\rm HC_3N}$ & ${\rm N_2}$ & CN & NH & HCN & ${\rm C_2H}$  & ${\rm NH_3}$ 
& ${\rm CH_3CN}$ & ${\rm CH_3OH}$ & ${\rm NH_2}$ & N & O & ${\rm H_2}$ & HCO & ${\rm C_3H_2}$    \\
${\rm CH_2OH}$ & & & & & & & & & & & & & & \\
\tableline
\end{tabular}}
%\end{tabular}
%\caption{\label{table1} Chemical reaction network used in the
%  calculation of the abundances of charged species.}
\end{center}
\end{table*}
%%%%%%%%%%%%%%%%%%%%%%%%%% TABLE 2 %%%%%%%%%%%%%%%%%%%%%%%%%%%%%%%%%%%%%

\begin{table}
\begin{center}
%{\large Table 1}\\
\caption{\label{table2} Initial Abundances}
\begin{tabular}{l l}
\tableline\tableline
${\rm H}$    & $1.00 \times 10^{-2}$     \\  
${\rm H_2}$  & $4.95 \times 10^{-1}$     \\
${\rm He}$   & $1.40 \times 10^{-1}$     \\  
${\rm C^+}$  & $7.30 \times 10^{-5}$     \\  
${\rm N}$    & $2.14 \times 10^{-5}$     \\  
${\rm O}$    & $1.76 \times 10^{-4}$     \\  
${\rm Si}$   & $2.00 \times 10^{-8}$     \\  
\tableline
\tableline
\end{tabular}
%\end{tabular}
%\caption{\label{table1} Chemical reaction network used in the
%  calculation of the abundances of charged species.}
\end{center}
\end{table}

\section{Chemical Network}\label{chemistry}

We use a subset of the UMIST chemical reaction network (Woodall et al.\ 2007), which includes data on 420 species and 13 elements followed in gas-phase reactions. 
Each chemical species is treated as a separate fluid in our formalism. The source and sink terms on the 
right-hand side of Eq.~(\ref{HDcontChem}) and in Eqs.~(\ref{thin13}) and (\ref{thin14}) describe the interactions 
between chemical species. They are given by: 
\begin{eqnarray}
S_j &=& \frac{\partial \rho_j}{\partial t} \nonumber \\
& = &\!\!\!\!\! \sum_{k\ne j, l \ne j} \!\!\!\!\! \alpha_{jkl} \rho_k \rho_l 
- \!\!\! \sum_{k\ne j, l} \!\!\!\! \alpha_{klj} \left( 1\!\!+\! \delta_{lj}\right)\rho_l \rho_j
+ \!\! \sum_{k\ne j} \!\! \left( \beta_{jk} \rho_k \!\! - \!\! \beta_{kj}\rho_j \right)\!, \nonumber \\
\end{eqnarray}
where $\alpha_{klj}$ is the rate coefficient for the production of species $j$ due to the interaction of 
species $k$ and $l$, $\beta_{jk}$ is the rate of production of species $j$ from species $k$ by 
photoreactions, and $\delta_{lj}$ is the Kronecker delta function.

In our models the source  terms are implemented in an operator-split fashion, and the change of abundances 
due to the source terms is updated after the advection terms.
The species we consider are listed in Table 1, and they include 78 gas-phase atomic and 
molecular species, and 33 grain ice mantle species, that are composed of the elements H, He, C, N, O, and 
Si as a representative low ionization potential metal. The chemical network follows 1553 reactions.

All the models studied in this work use the isothermal approximation (small deviations from isothermality have been shown to not affect the dynamical evolution of cores significantly, e.g. Keto \& Caselli 2010). We adopt as a fiducial temperature  
$T=10$K, consistent with line intensity ratios of different molecular transitions showing that starless 
cores are almost isothermal, with $T\sim 10$ K, (Tafalla et al.\ 1998, 2002, 2004), although there may be 
small variations of the order of a few K (e.g., Evans et al.\ 2001). In our parameter studies, we test for the effect of small temperature variations on our chemical abundances.

Since the prestellar cores we model are embedded in a molecular cloud we assume a visual extinction 
$A_{\rm V}=3$ mag at the outer boundary, so that the results are not much affected by photodissociation. 
For the inner layers of the core we obtain the visual extinction from the column density to the surface 
using the empirical conversion relation $N_{\rm H_2}/A_{\rm V}=9.4 \times 10^{20}$ ${\rm cm^{-2}mag^{-1}}$ 
(Bohlin et al.\ 1978). For our deeply embedded cores, no ${\rm H_2}$ self-shielding is required;  CO self-shielding on the other hand is accounted for based on the model of Lee et al.\ (1996). 

We assume  a standard MRN distribution of grains (Mathis, Rumpl \& Nordsieck 1977),
assumed to be 
well-mixed with the gas. Since most of the grains are negatively charged at the densities of interest 
in prestellar cores, we also take into account the electrostatic attraction between ions and grains 
through an enhanced rate of accretion for the ions (Rawlings et al.\ 1992). 
We use a fixed grain abundance ($n_{\rm grain}/n_{\rm neutral} = 10^{-12}$) and charge 
(we take all grains to be singly negatively charged). Grain processes modeled include freeze-out, 
surface reactions, thermal desorption, and cosmic ray heating. We have used a sticking probability of 0.3. 

The discrete nature of grains means that chemistry on their surfaces should properly be modeled by 
stochastic methods (e.g.; Stantcheva, Shematovich, \& Herbst 2002; Chang et al.\ 2007). However, 
those methods are computationally very expensive, especially for a large chemical network and the 
large number of HD and MHD models we run. Instead, we use the rate equations method, with modifications as 
suggested by Vasyunin et al. (2009), which brings the results of this method into good agreement 
with Monte Carlo modeling for the range of temperatures we are concerned with. The grain reaction 
network is taken from Hasegawa \& Herbst (1993), as are the rates for desorption by cosmic ray 
heating of grains (updated for new binding energies where necessary). 
Where available, we use binding energies determined in the lab (see Willacy 2007). 
Other binding energies are taken from Hasegawa \& Herbst (1993). The ${\rm CO_2}$ chemistry on the grains is based on Garrod \& Pauly (2011)\footnote{ with the activation barrier for the ${\rm CO+O \rightarrow CO_2}$ reaction on the grain ice increased from 290 K to 340 K (within the experimental range).}.

The gas phase reactions are better constrained as they have been studied for a longer time, and there 
appears to be convergence in the literature among different groups (e.g., Aikawa et al.\ 2003) with the exception of sulphur-bearing molecules (Wakelam, Herbst \& Selsis 2006), which, however, are not considered in our model. In our study we aim to assess the effect of the dynamics on chemical abundances, while the chemical network remains fixed.

We use low-metal initial abundances\footnote{with the exception of Si, the abundance of which is higher than in the low-metal model}, given in Table 2. The calculations are initiated with all elements other than molecular hydrogen in the form of atoms. Of course in nature there is some chemical evolution taking place during the formation 
of the parent cloud.  A first attempt to calculate initial abundances for molecular 
clouds for one particular cloud formation scenario was presented by Hassel et al. (2010). However, these 
results are model-dependent and will differ for different cloud formation models. We will return to the 
issue of  initial abundances in a future publication.

\section{Parameter Study}\label{parstu}

For each class of models (non-magnetic or magnetic), we examine the dependence of the molecular abundances and their evolution on four parameters:  the C/O ratio, the cosmic ray ionization rate, the temperature, and a parameter controlling the time available for chemical evolution. In non-magnetic models, the latter is a time interval prior to collapse during which chemistry evolves but the core is assumed to not evolve dynamically. In magnetic models, it is the initial mass-to-magnetic-flux ratio which controls the amount of magnetic support: a smaller initial mass to magnetic flux ratio implies a larger amount of magnetic support against gravity, and hence a slower dynamical evolution for the core. 

Our ``reference'' non-magnetic model has a C/O ratio of 0.4, a cosmic ray ionization rate of  
$\zeta = 1.3 \times 10^{-17} {\rm \, s^{-1}}$, a temperature $T=10 {\rm \, K}$ and an initial collapse delay time of 1 Myr. Our ``reference'' magnetic model similarly has a C/O ratio of 0.4, $\zeta = 1.3 \times 10^{-17} {\rm \, s^{-1}}$, $T=10 {\rm \, K}$, and a central mass-to-flux ratio equal to the critical value for collapse (Mouschovias \& Spitzer 1976).

For non-magnetic models, we examine two additional values of delay: zero, and 10 Myr. For magnetic models, 
we examine two additional values of the initial mass to magnetic flux ratio: 1.3 times the critical value 
(a faster-evolving, magnetically supercritical model), and 0.7 of the critical value (a slower, 
magnetically subcritical model).
The three values of the delay and the three values of the mass-to-flux ratio set our dynamical model variations.
 When these models are combined with reference values for the C/O ratio, $\zeta$, and $T$, we will refer to them as the six basic dynamical models.

The carbon-to-oxygen ratio is varied from its reference value by keeping the abundance of C constant 
and changing that of O (as in Terzieva \& Herbst 1998). The two other values of C/O ratio examined are 1 and 1.2. We have varied each of the six basic dynamical models by changing its C/O ratio to one of the values above, and we have thus studied 12 models (6 magnetic and 6 non-magnetic) with the C/O ratio varied from its reference value.

 We have also considered different temperatures for each of the six basic dynamical models.   We have varied the temperature by a factor of $\sim 1.5$ from its reference value of $10$ K and examined models with $T=7$ K and $T=15$ K, and we have thus studied 12 models (6 magnetic and 6 non-magnetic) with a temperature differing from its reference value.  

Finally, to test the effect of the cosmic ray ionization rate we have studied two additional 
magnetic and two non-magnetic models, which are, in each case, identical to the corresponding 
``reference'' models, but with different cosmic-ray ionization rates: namely, a factor of four above ($\zeta = 5.2 \times 10^{-17}$ $s^{-1}$) and below 
($\zeta = 3.3 \times 10^{-18}$ $s^{-1}$) its reference value (covering the range of observational 
estimates e.g.: MaCall et al.\ 2003; Hezareh et al.\ 2008). 

Thus, we have examined a total of 34 models (6 basic dynamical models, 12 with varied C/O ratios, 12 with varied $T$, and 4 with varied $\zeta$).  These 34 models are depicted in our figures in \S \ref{res} with different line types, thicknesses, colors, and symbols, as follows:
\begin{itemize}
\item Line type (solid, dashed, dotted) and symbols (X, diamond, cross) depict the evolutionary timescale (``reference'', ``slow'', and ``fast'' respectively). 
\item Color denotes the type of model (magnetic or non-magnetic), with bluish hues corresponding to non-magnetic models, and reddish hues corresponding to magnetic models.
\item The line thickness/symbol size (for red and blue lines and symbols) indicate the values of the cosmic ray ionization rates: thin red/blue solid lines and small red/blue symbols correspond to the low value of $\zeta$, while thick red/blue solid lines and large red/blue symbols to the high value of $\zeta$, with all other parameters fixed at their ``reference'' values. 
\item The effect of different C/O ratio values is shown as a shaded area (orange for magnetic runs, and cyan for non-magnetic runs). The edge of the shaded area in contact with the corresponding (``fast'', ``reference'', or ``slow'', magnetic or non-magnetic) model line corresponds to C/O=0.4, while the edge of the shaded area furthest away from the line corresponds to C/O=1.2
\item Finally, the effect of the temperature is shown with the colors brown for magnetic runs and purple for non-magnetic runs, either as a shaded area, or as line thickness/symbol size. In the latter case, the thick brown/purple lines (or the large brown/purple symbols) correspond to $T=15$ K and the thin brown/purple lines (small brown/purple symbols) to $T=7$ K.  
\end{itemize}

\section{Results}\label{res}

\subsection{Dynamics}

\begin{figure}
\plotone{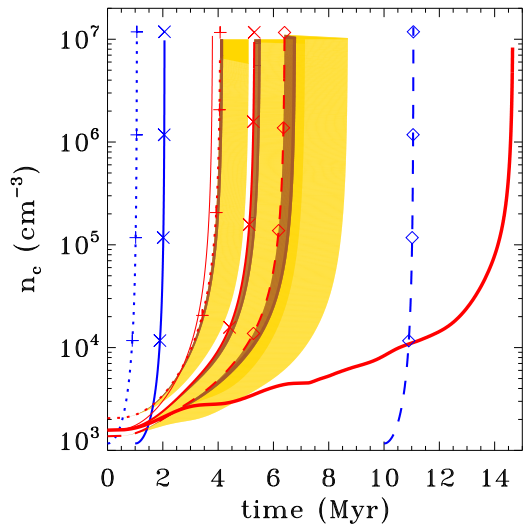}
\caption{\label{n_vs_t} 
(a) Central number density, $n_{\rm c}$,  as a function of time.  
Symbols indicate the times at which full radial snapshots are taken. 
Blue lines: non-magnetic models (dotted/crosses: no delay; solid/x: 1 Myr delay; 
dashed/diamonds: 10 Myr delay). Red lines: magnetic models (dotted/crosses: supercritical model; 
solid/x: critical model; dashed/diamonds: subcritical model). 
The thin and thick red solid lines correspond to cosmic 
ray ionization rate $\zeta$ a factor of four above and below the ``reference'' value, 
and bracket the normal-thickness red solid line. 
The yellow shaded areas show the effect of a varying C/O ratio, 
and correspond to a range of C/O values between $0.4$ (reference value) and $1.2$.
The brown shaded areas show the effect of varying the temperature from 7K to 15K, 
the reference value being 10K.}
\end{figure}

\begin{figure*}
\plotone{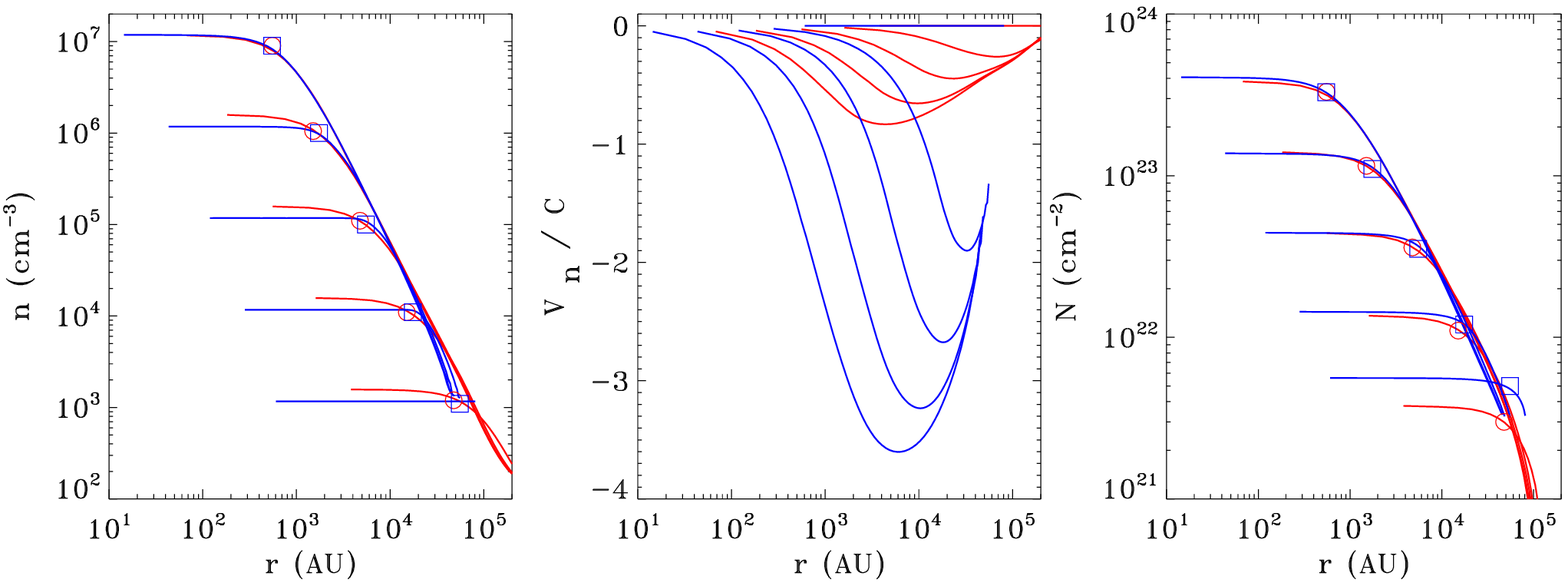}
\caption{\label{comp_dynamics_rad} 
Radial profiles of: the number volume density, $n$ (panel a); velocities of the neutrals in units 
of the isothermal sound speed (panel b); and (edge-on for the magnetic
runs) column density (panel c). Red lines: magnetic runs; 
Blue lines: non-magnetic runs. Each curve corresponds to a snapshot in time when the volume density is an 
order of magnitude higher than the previous one. Time elapsed per snapshot in magnetic runs: 
3.53 Myr, 4.4 Myr, 4.6 Myr, 4.65 Myr; in non-magnetic runs: 1.88 Myr, 2.01 Myr, 2.05 Myr, 2.06 Myr.}
\end{figure*}

Figure \ref{n_vs_t} shows the evolution of the central density of contracting core models with time. 
Red lines correspond to magnetic runs, and blue lines to non-magnetic runs. The different magnetic runs 
correspond to different values of the mass-to-magnetic-flux ratio: critical for the solid lines/ x symbols; 
0.7 times the critical value (magnetically subcritical, slower-evolving magnetic model) for the dashed 
lines/diamonds; and 1.3 times the critical value (magnetically supercritical, 
faster-evolving magnetic model) for the dotted lines/ + symbols. The different non-magnetic runs 
correspond to different assumed delay times, during which the chemistry operates, before the core is allowed 
to evolve dynamically (collapse): 1 Myr delay for the solid lines/ x symbols; 10 Myr delay 
(slowest-evolving non-magnetic model) for the dashed lines/diamonds; and no delay (fastest-evolving 
non-magnetic model) for the dotted lines/ x symbols. The symbols mark the times when full radial outputs
are obtained and analyzed, and correspond to successive enhancements of the central density by an order
of magnitude.

Different dynamical models differ with respect to the two fundamental ingredients affecting the chemical 
evolution of a core: density, and time spent at each density. Non-magnetic models 
with substantial time delays (representing the time it takes for some level of turbulent support to 
dissipate) can take as much time as substantially magnetically subcritical models to reach very high 
densities. However, once the collapse sets in (support dissipates), the evolution occurs essentially within one 
free-fall timescale, so that the bulk of the time delay is spent at the initial, low density ($10^3 {\rm cm ^{-3}}$). In contrast, 
slow magnetic models allow the core to spend substantial time at {\em higher} densities, and as a result 
the chemical state of these models are expected to be different.  

 The thin and thick red lines correspond to
magnetically critical models but have $\zeta$ a factor of four above and below the 
``reference'' value, and they bracket the normal-thickness red solid line. Because the ambipolar diffusion timescale depends on the degree of ionization, which in turn depends on the cosmic-ray ionization rate, 
$\zeta$ affects the evolutionary timescale of the magnetic models: a higher degree of ionization implies 
increased coupling between magnetic field and neutral fluid, and a resulting slower evolution, as 
magnetic forces are more effective in opposing gravity (see Paper II). 

In addition to the mass-to-flux ratio and the degree of ionization, the evolutionary timescale of the magnetic models is also affected by the C/O ratio. The effect of different values of the C/O ratio can be seen in Fig.~\ref{n_vs_t} as the orange shaded area, which denotes a range of C/O values between 0.4 (reference) and 1.2. The reason for this dependence is that the C/O value affects the chemical reaction network and consequently the degree of ionization at each density, resulting to a different evolution rate. In general, higher values of the C/O ratio result in slower evolution, and the effect is more pronounced for models with a low mass-to-flux ratio. 

Finally, the evolutionary timescale of the magnetic models also depends on the temperature. The effect of different $T$ values is shown in Fig.~\ref{n_vs_t} as the brown shaded area, which denotes a range of $T$ values between $7$ K and $15$ K. The dependence is stronger for slower models, and it enters through the dependence of the degree of ionization on temperature (see Paper II). The evolutionary timescale increases with increasing temperature, and the difference between $10$ K and $15$ K is significantly larger than the difference between $7$ K and $10$ K, although the fractional increase is comparable (about 50\% in each case). 

Radial profiles of the density, velocity, and (edge-on for the
magnetic runs) column density for the magnetic and non-magnetic reference 
models and for the different time snapshots corresponding to the x symbols in Fig.~\ref{n_vs_t} are given
in Fig.~\ref{comp_dynamics_rad}. Again the magnetic model is represented by red lines and the non-magnetic 
model by blue lines. Each snapshot corresponds to a time when the volume density is an order of magnitude 
higher than the previous one. The time elapsed between the initialization of each simulation and each 
snapshot for the magnetic model is 3.53 Myr, 4.4 Myr, 4.6 Myr, 4.65 Myr. The corresponding times for the 
non-magnetic model are 1.89 Myr, 2.02 Myr, 2.05 Myr, 2.06 Myr. The first snapshot in both cases corresponds 
to the initial configuration of each model. 

Radial profiles of the volume density are shown in the left panel. These are power-law profiles with a 
flat inner region, the boundary of which is indicated with a circle for the magnetic runs and a square 
for the non-magnetic runs. Profiles of the infall velocity, in units of the sound speed, are shown in 
the middle panel. Here the difference between the magnetic and the non-magnetic models is most striking, 
as for the magnetic model the infall is always subsonic, while for the non-magnetic model the infall 
velocities are substantially supersonic throughout the evolution of the core and for most of its radial 
extent. The right panel shows radial profiles of the column density, which have similar properties with 
the volume density profile. 

The masses of the different model cores are not exactly the same. This is because the mass of each core depends on the temperature  in the case of non-magnetic cores, and the mass-to-flux ratio, temperature, C/O ratio, and cosmic-ray ionization rate in the case of magnetic cores. In spite of these dependencies, we have selected the initial conditions of our models so that the core masses remain roughly comparable: the non-magnetic ``reference'' model core mass is $15 {\rm \, M_\odot}$ and the magnetic ``reference'' model core mass is $20 {\rm \, M_\odot}$, while in general the core masses in each class of models do not differ from this value by more than a factor of $\sim 2$.

A shared feature of the volume and column density radial profiles with important observational 
implications is the size of the flat-profile inner region. It is well-known that prestellar 
cores have uniform column density inner regions (Ward-Thompson et al.~1994; Ward-Thompson et al.~1999; 
Bacmann et al.~2000; Shirley et al.~2000; Schnee et al.~2010). The angular size of the 
{\em column density} profile flat inner region, which is observationally measurable in well resolved cores,
combined with an estimate for the distance can yield the physical size of the column density profile 
flat inner region, which is of the same size as the {\em volume density profile flat inner region}, 
as can be directly seen by comparing the left and the right panels of Fig.~\ref{comp_dynamics_rad}. 
The latter can be used to estimate the central volume density of a collapsing core  (Tassis \& Yorke 2011).

\begin{figure*}
\plotone{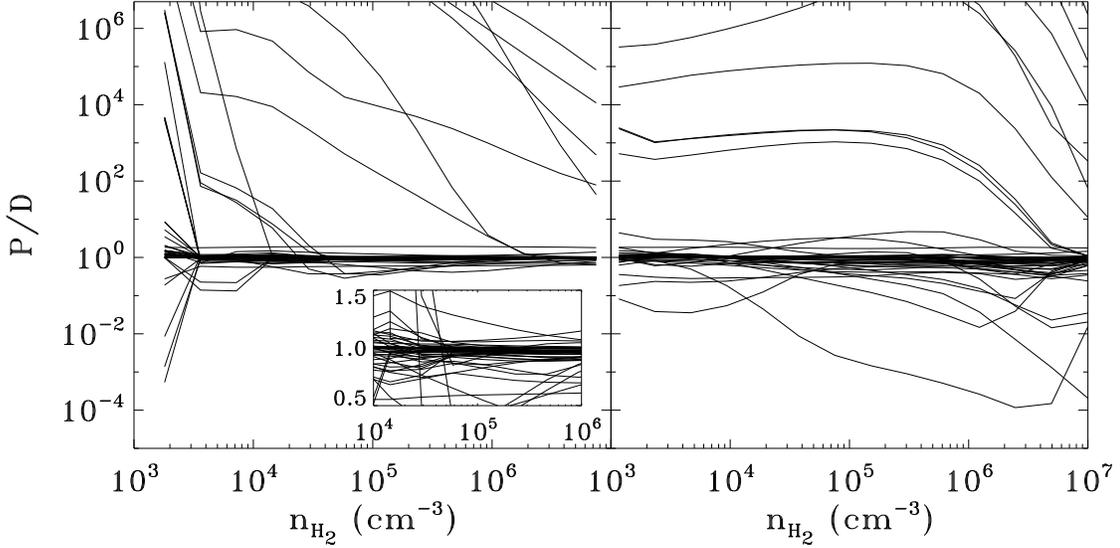}
\caption{\label{NEQ} Ratio of total (including all possible channels) production and destruction rates as 
a function of central density, for the magnetic (left panel) and the non-magnetic (right panel) reference 
models. The inset in the left panel shows a zoom-in of the density range between $10^4$ and 
$10^6$ $cm^{-3}$, with a linear y axis between $P/D=0.5$ and $1.5$.
 }
\end{figure*}

\subsection{Non-equilibrium Chemistry, Production and Destruction of Commonly Studied Molecules}

%%%%%%%%%%%%%%%%%%%%%%%%%%%%%%%%%%%%%%%%%%%%%%%%%%%%%%%%%%%%%%%%%%%%%%%%
\begin{figure*}
\plotone{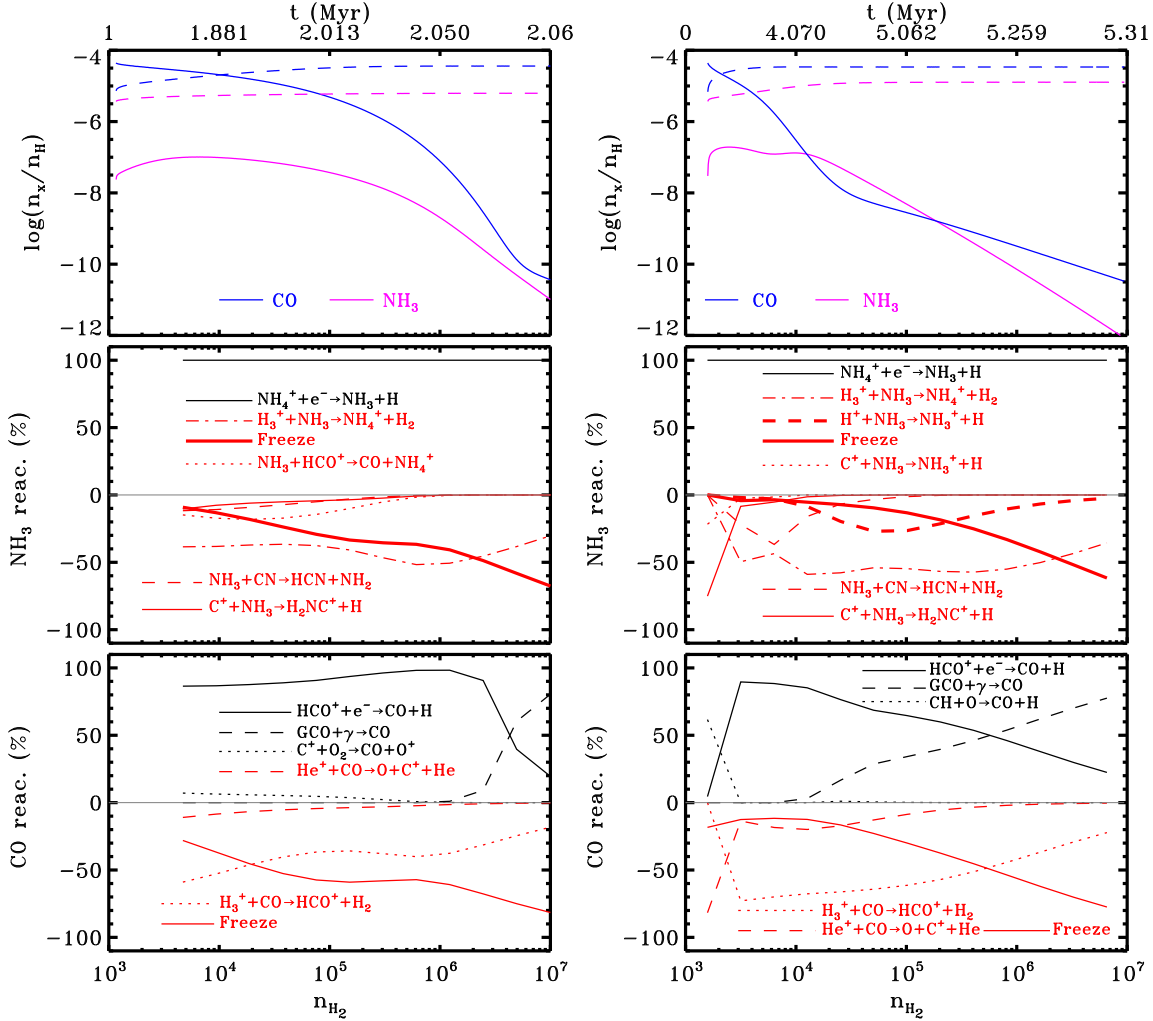}
\caption{\label{reacfig} 
 Top panels: central abundances for two commonly observed molecules (CO and NH$_3$), for the  
non-magnetic (left column) and magnetic (right column) reference models. Solid lines: gas phase abundance; 
dashed lines: grain ice mantle abundance. Middle and bottom panels: 
percentage contribution of most important reactions to the total production (black lines) and 
destruction (red lines) rates for the two molecules.}
\end{figure*}

%%%%%%%%%%%%%%%%%%%%%%%%%%%%%%%%%%%%%%%%%%%%%%%%%%%%%%%%%%%%%%%%%%%%%%%%
The need for the inclusion of non-equilibrium chemistry in our dynamical models is explicitly demonstrated 
in Fig. \ref{NEQ}. Here we plot the ratio $P/D$ between the total production and destruction rates of 
various molecules, as a function of density, at the center of our reference core models. The left panel 
corresponds to the magnetic reference model, and the right panel to the non-magnetic reference model. The 
total production and destruction rates are defined as including all chemical processes leading to the 
production or destruction of a specific molecule respectively. Each line corresponds to a different 
molecule. Note that the y-axis in the two plots is logarithmic and extends over ten orders of magnitude. 
A large number of molecules are very much out of equilibrium ($P/D$ is very far from 1), and they remain 
so throughout each simulation. In the magnetic run, where more time is available for the system to 
approach chemical equilibrium even at high densities, the majority of molecules eventually move closer 
to $P/D =1$; in the non-magnetic model the reverse trend is observed, and as the density increases and 
the evolution timescale decreases, most molecules deviate further away from $P/D = 1$. However, even for the 
magnetic model the spread about $P/D=1$ is substantial, as demonstrated by the inset on the left panel, 
which zooms in around central densities of $10^4 - 10^6$ particles per cm$^3$ and $P/D$ within 50\% of 
unity displayed on a linear scale.  In the inset, the values of $P/D$ exhibit a spread of more than 20\%  
around unity. 

The effect of non-equilibrium chemistry becomes obvious when we examine the dependence of central molecular 
abundances on central density for different dynamical models. If the chemistry was in equilibrium, then 
the central volume density would uniquely determine the central molecular abundances. This, however, is 
not the case, as we can see in Fig. \ref{reacfig}, where we examine the central abundances, production, 
and destruction rates due to various reactions for one carbon-bearing (CO) and one nitrogen-bearing 
(NH$_3$) molecule. The left column corresponds to the non-magnetic reference model (1 Myr delay), and the 
right column corresponds to the magnetic reference model (critical mass-to-flux ratio). The top row shows 
the evolution with central density (bottom axis) and with time (top axis) of the central abundances of 
CO (blue lines) and NH$_3$ (purple lines). Solid lines correspond to gas-phase abundances and dashed 
lines correspond to abundances on grain ice mantles. The middle and bottom rows show the most important 
production (black lines) and destruction (red lines) reactions for the two molecules. The importance of 
each reaction is quantified by the percentage by which the specific reaction contributes to the total 
production/destruction rate of each molecule (y-axis). 

Comparing the top left and right panels, we can immediately see that the central abundances of the two 
molecules at the same central density are very different for different dynamical models, a 
result of the chemistry being out of equilibrium. Because each model spends different amounts 
of time at each density (see Fig. \ref{n_vs_t}), non-equilibrium chemistry results to significantly 
different central abundances. Note that since the magnetic model evolves more slowly, the gas phase 
abundances of both molecules in the magnetic core model decrease more steeply with density, as the 
molecules freeze out on the grain mantles; correspondingly, there is a faster increase in the 
ice mantle abundance of each molecule. 

It is evident from the middle and bottom panels that any specific reaction reaches its maximum importance 
at different densities for different dynamical models. Freeze out onto grain mantles (solid red line), 
which becomes increasingly important at higher densities, is a prominent example. 
Furthermore, for each molecule different reactions dominate the production and destruction 
channels for different dynamical models.

\subsection{Evolution of central abundances}\label{evca}

%%%%%%%%%%%%%%%%%%%%%%%%%%%%%%%%%%%%%%%%%%%%%%%%%%%%%%%%%%%%%%%%%%%%%%%%

\begin{figure*}
\plotone{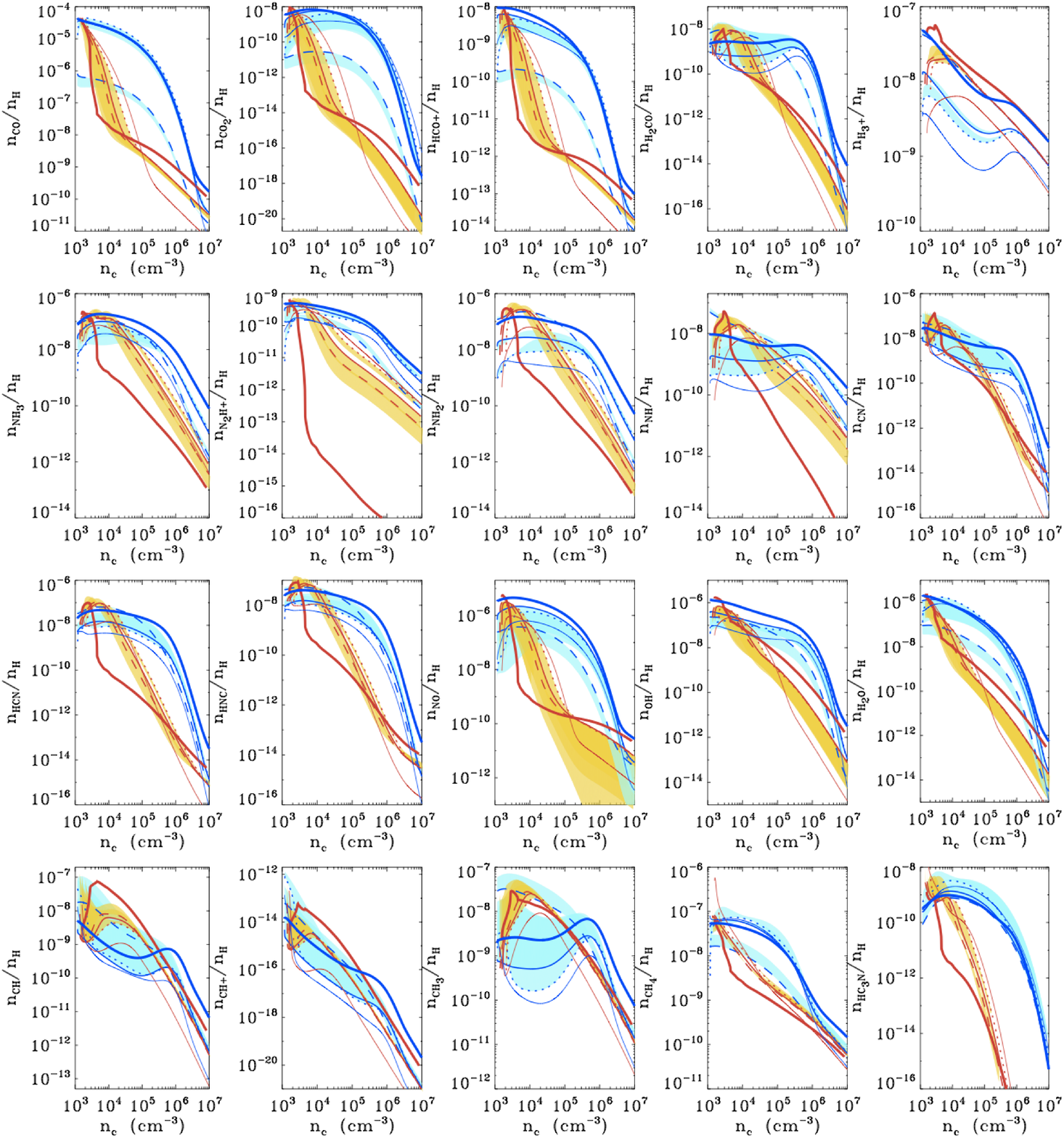}
\caption{\label{comp_cen_abund} 
Central abundances for representative molecules as a function of central density, for different dynamical 
models, different C/O ratios, and different cosmic ray ionization rates. Reddish hues correspond to magnetic 
models (dashed: magnetically subcritical; solid: magnetically critical; dotted: magnetically supercritical), 
and bluish hues to non-magnetic models (dashed: 10 Myr delay; solid: 1 Myr delay; dotted: no delay). 
The thin and thick solid lines correspond to cosmic 
ray ionization rate $\zeta$ a factor of four above and below the ``reference'' value. The orange/cyan shaded areas show the effect of a varying C/O ratio in magnetic/non-magnetic models and correspond to a range of C/O values between $0.4$ (reference value) and $1.2$. }
\end{figure*}

\begin{figure*}
\plotone{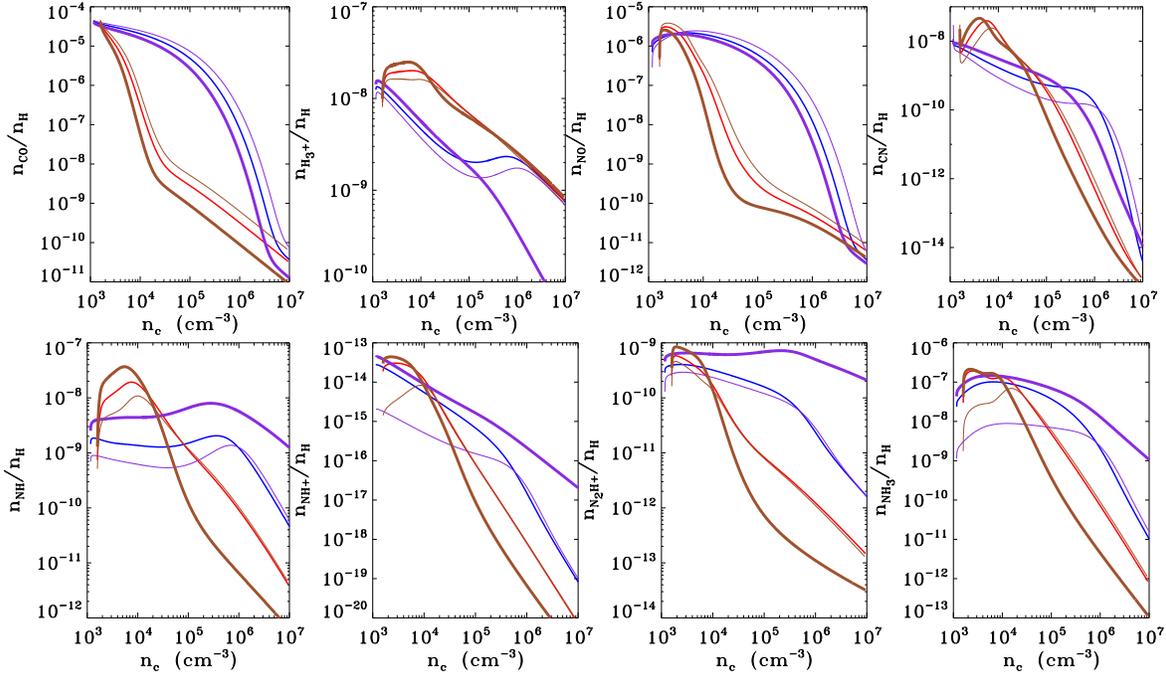}
\caption{\label{T_cen_abund} 
Effect of temperature on central abundances of representative molecules. Blue/red lines: reference non-magnetic/magnetic models.  Thick/thin brown lines: $T=15$ K/ $T=7$ K magnetic models; all other parameters kept at their reference values. Thick/thin purple lines: $T=15$ K/ $T=7$ K non-magnetic models; all other parameters kept at their reference values.  }
\end{figure*}

\begin{figure}
\plotone{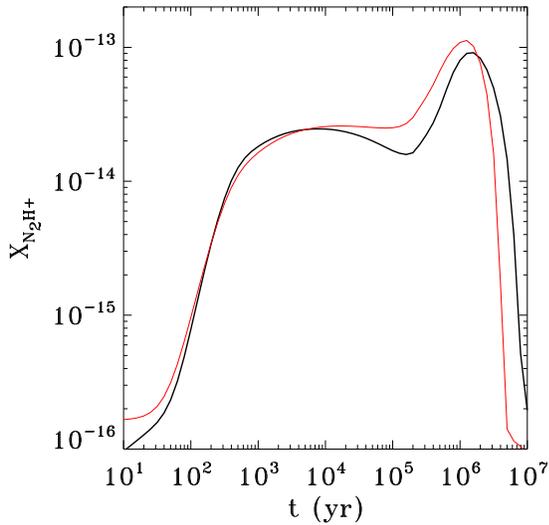}
\caption{\label{twostatic} 
Evolution of N$_2$H$^+$ abundance for $T=10$ K (black line)  and $T=15$ K (red line) in a static medium of density $2\times 10^4 {\rm , cm^{-3}}$.  }
\end{figure}
%%%%%%%%%%%%%%%%%%%%%%%%%%%%%%%%%%%%%%%%%%%%%%%%%%%%%%%%%%%%%%%%%%%%%%%%%%%%%%%%%%%%%%
\begin{figure*}
\plotone{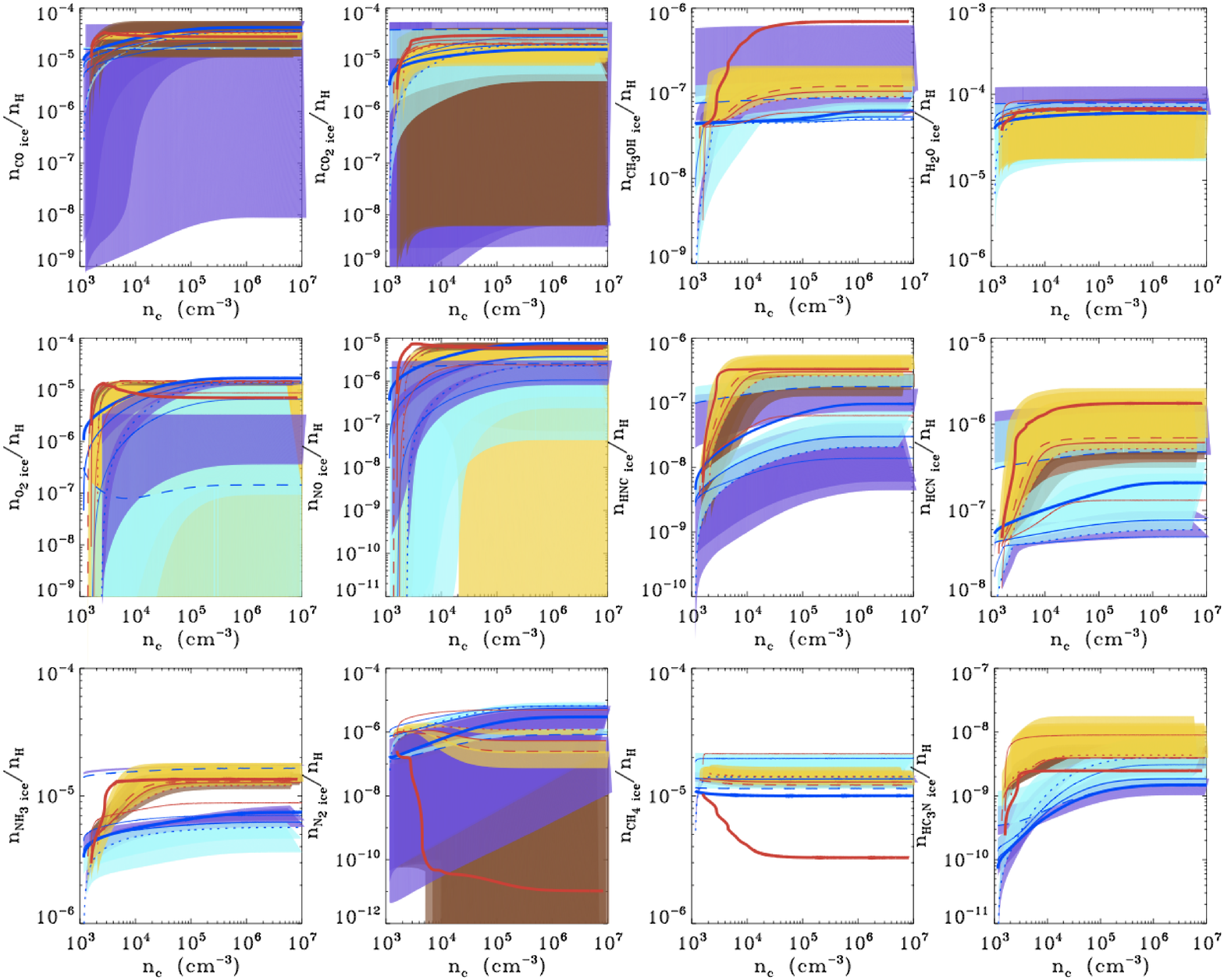}
\caption{\label{gmices} 
As in Fig.~\ref{comp_cen_abund}, for grain mantle ices. 
 Reddish hues correspond to magnetic 
models (dashed: magnetically subcritical; solid: magnetically critical; dotted: magnetically supercritical), 
and bluish hues to non-magnetic models (dashed: 10 Myr delay; solid: 1 Myr delay; dotted: no delay). 
The thin and thick solid lines correspond to cosmic 
ray ionization rate $\zeta$ a factor of four above and below the ``reference'' value. The orange/cyan shaded areas show the effect of a varying C/O ratio for magnetic/non-magnetic models and correspond to a range of C/O values between $0.4$ (reference value) and $1.2$. The brown/purple shaded areas show the effect of varying temperature for magnetic/non-magnetic models and correspond to a range of $T$ values between $7$ K and $15$ K (reference value is $10$ K). }
\end{figure*}

Figures \ref{comp_cen_abund} and \ref{T_cen_abund}  show the evolution with central density of the central abundances of selected molecules in the gas phase. Figure \ref{comp_cen_abund} shows the effect of the dynamics, C/O ratio, and $\zeta$ on the gas-phase abundances, while Fig.~ \ref{T_cen_abund} shows the effect of the temperature. 

In Fig.~\ref{comp_cen_abund}, reddish hues correspond to 
magnetic models and bluish hues correspond to non-magnetic models. Different line types correspond to different dynamical 
models, with dashed lines representing the ``slowest'' models (subcritical magnetic model and 10 Myr delay 
non-magnetic model); solid lines representing the ``intermediate'' models (critical magnetic model and 
1 Myr delay non-magnetic model); and dotted lines representing the ``fast'' models (supercritical magnetic 
model and no-delay non-magnetic model). 
The shaded bands (cyan for non-magnetic models, orange for magnetic models) show the effect of the C/O ratio on the abundances (the edge of the band adjacent to a model line corresponds to C/O=0.4 and the edge furthest away from the line to C/O=1.2). The line thickness is representative of the value of the cosmic ray ionization rate, with normal thickness lines corresponding to $\zeta =  1.3 \times 10^{-17} {\rm \, s^{-1}}$, and thick (thin) lines corresponding to values of $\zeta$ a factor of four above (below) that value. 

A general feature of the central gas-phase abundance evolution curves of Fig.~\ref{comp_cen_abund} is their 
dependence on the dynamical evolution model. Magnetic models, which evolve more slowly between different 
central densities and spend more time at each value of the density, have an abundance peak at low 
densities, and generally exhibit a faster drop of the gas-phase abundance with increasing central density, 
due to efficient freeze-out onto grain ice mantles. As a result, the abundance evolution curves appear convex. 
In contrast, non-magnetic models evolve faster,  
chemistry has less time in which to operate at each density, the abundances peak and drop at higher densities, 
and the non-magnetic abundance evolution curves are generally concave. The quantitative explanation of this behavior can be given by comparing the dependence of the freeze-out timescale and the collapse timescale on density for different models. The steep drop in molecular abundances (especially in carbon-bearing molecules) occurs at the density where the freeze-out timescale becomes shorter than the collapse timescale (Walmsley 1991). In magnetic models this happens at much lower densities ($\sim 10^4 {\rm \, cm^{-3}}$) than in non-magnetic models ($\sim 10^6 {\rm \, cm^{-3}}$). 

The results of Fig.~\ref{comp_cen_abund} make it obvious that different collapse delay times at the 
initial phases of non-magnetic core evolution are not a good proxy for the effect of magnetic fields. 
Although the slowest non-magnetic model curves (dashed curves, 10 Myr delay) are generally closer to the 
magnetic model curves, the difference is still significant even in the qualitative behavior of the abundances 
(concave vs convex). The reason for this is that allowing more time for 
chemical evolution at $\sim 10^3 {\rm \, cm^{-3}}$ does not have the same effect, from a chemistry 
perspective, as spending a similar  amount of time at higher densities, which is what happens in magnetic 
models. 

The behavior of certain molecules, such as H$_2$CO, H$_3^+$, NH,  and CH$_3$, is quite different for the non-magnetic models. While the slowest 
non-magnetic and all the magnetic models have a qualitative similar behavior that resembles that of 
other molecules  (the abundance builds up with increasing density, peaks at $\sim 10^4 {\rm \, cm^{-3}}$, 
and then decreases at even higher density),  the faster non-magnetic models, instead of a peak, 
exhibit a local minimum at these densities. 

The molecules with the greatest sensitivity to the  C/O ratio are, unsurprisingly, the oxygen-bearing 
molecules, the abundance of which is generally lower for higher C/O ratios. Molecules with more than one oxygen atoms are affected more than molecules with a single oxygen atom (see, e.g., CO versus CO$_2$). Other molecules sensitive to the C/O ratio include CH, CH$^+$, CH$_3$, and NH. 

As expected, H$_3^+$ responds directly to the change in the cosmic-ray ionization rate, as it is a direct ionization product (see Paper II for more details). Molecules that are primarily produced through reactions involving direct ionization products, such as H$_3^+$ and He$^+$ (e.g., CH$^+$) or free electrons, (e.g., NH$_2$, NH, and CH$_3$) are similarly affected, even in the non-magnetic models. The magnetic models have an additional response to $\zeta$ through the modification of their evolutionary timescale (see Fig.~\ref{n_vs_t}). For this reason, the change in molecular abundances, especially in the case of the largest-$\zeta$ magnetic model (which is also by far the slowest-evolving model) can be dramatic and feature very significant depletion in relatively low densities. The biggest difference is seen in molecules which, for moderate evolutionary timescales, show relatively modest depletion (e.g. N-bearing molecules).

Figure \ref{T_cen_abund} shows the effect of the temperature on the evolution with density of the central abundances of a few representative molecules. The blue and red lines correspond to the reference non-magnetic and magnetic models (the same ones shown with the normal-thickness blue and red lines in Fig.~\ref{comp_cen_abund}). High-temperature ($T=15$ K) models are shown with the thick brown and purple lines, for magnetic and non-magnetic models respectively, while low-temperature ($T=7$ K) models are shown with the thin brown and purple lines; all other parameters are kept at their reference values. 

There are some significant differences between the
molecular abundances in the 15K models compared
to those in the 10K models, especially for nitrogen-bearing 
compounds.  This is a result of two effects: 
(1) the reaction
\begin{equation}\label{react:nhplus}
\hbox{N}^+ + \hbox{H}_2 \longrightarrow \hbox{NH}^+ + \hbox{H} 
\end{equation}
has a small activation barrier of 85 K.  At 15 K the temperature
is high enough for this barrier to be overcome and hence this 
reaction is much more efficient than at 10K (see also
Wakelam et al. 2010).  This reaction is the start of the
formation of nitrogen-bearing molecules, and an increase in 
its efficiency results in higher abundances of these molecules.
(2) At 15 K the grains are warm enough for the desorption of N$_2$
ice to proceed more rapidly than at 10K (the rate of desorption
is approximately 10$^{10}$ times greater at 15 K compared to 10K).  

The way in which these processes act to produce the model results
depends on the conditions in the models.  As can be seen in 
Figure  \ref{T_cen_abund}, increasing the temperature to 15K results in an increase
in the abundance of nitrogen bearing molecules for the non-magnetic models
(bold purple curves).
The opposite effect is seen for the magnetic models (bold brown curves).
The difference between the two types of models is due to the timescales
over which they evolve.  The non-magnetic models in Fig.  \ref{T_cen_abund}  evolve quickest and
reach their final density within 2 Myr, whereas the magnetic models
take between 5.3 and 5.5 Myr (Fig.~\ref{n_vs_t}).  Taking the non-magnetic models first:
there is an increase in gaseous N$_2$ caused by its faster desorption from the grains (see Fig.~\ref{gmices}). 
This means that the N$_2$ abundance in the gas is higher
at earlier times.  The increase in N$_2$ increases the destruction
rate of H$_3^+$ as they react forming N$_2$H$^+$, and
results in the sharp decrease in H$_3^+$ abundance seen in Fig.~\ref{T_cen_abund}.
Most of the N$_2$H$^+$ will be recycled back into N$_2$ by recombination with
electrons, but a small amount ($\sim$ 5\%; Molek et al.~2009) will form
NH.  This can go on to form other gaseous molecules, or it can freeze out
onto the grains, where it is quickly hydrogenated to form ammonia ice. 
Ammonia has a much higher binding energy than N$_2$ and once formed
remains on the grains at these temperatures.  The effect of the
enhanced rate for reaction~(\ref{react:nhplus}) can also be seen in the
increased abundance of NH in the gas phase in the 15 K model.

The magnetic models are indicated by red/brown lines in Fig.~\ref{T_cen_abund},
where the bold brown line is the 15 K model.  The increased temperature 
here has exactly the opposite effect on nitrogen molecular abundances
compared to the non-magnetic models.  In this case the
abundance of N$_2$H$^+$ falls at 15 K, as do the abundances of
most other nitrogen-bearing species.  The abundance of H$_3^+$ is
roughly the same for all temperature models.  The reason
for this variance is the differing dynamical timescales of the
models considered.  The magnetic models remain at a relatively low density
for $>$ 2 Myr, before the onset of rapid collapse.  During this
prolonged period of time the chemistry evolves quite differently
depending on the temperature.  This is illustrated in Fig.~\ref{twostatic} which
shows the evolution of the N$_2$H$^+$ abundance with time for two static
models with a density of 2 $\times$ 10$^4$ cm$^{-3}$ and with
temperatures of 10 K (black) and 15 K (red).  At 15 K the
nitrogen chemistry evolves faster, with X(N$_2$H$^+$) reaching a peak value
of $\gtrsim 10^{-13} {\rm \, cm^{-3}}$ at 1 Myr, compared to $9\times 10^{-14} {\rm \, cm^{-3}}$ 
at almost 2 Myr for the 10 K model.
In the magnetic models, rapid collapse does not begin
until after $\sim$4 Myr.  From Fig.~\ref{twostatic}  we can see that the N$_2$H$^+$ abundance
is falling rapidly at this time and is more than an order of magnitude higher at 10 K than at 15 K. 
At 15 K much more nitrogen has been incorporated into NH$_3$ ices by this
stage in magnetic models (see Fig.~\ref{gmices}) and less is available in the gas phase.
Hence the early evolution of  the collapse is quite different and
results in the different abundances seen in Fig.~\ref{T_cen_abund}.

Figure \ref{gmices} shows the behavior of central abundances of various molecules in grain ice mantles, 
as a function of central density. Colors and lines are as in Fig.~\ref{comp_cen_abund},  however here the effect of temperature is also overplotted with a brown shaded area for the magnetic models, and a purple shaded area for the non-magnetic models. In all models, the composition of ice mantles is dominated by H$_2$O, in accordance with observations of absorption features in the spectra of background stars viewed through dense clouds taken in IR (e.g., Whittet 2003 and references therein). In observed dark, quiescent regions (where cores are most likely to be prestellar and the effect of the protostellar UV field is minimal), CO and CO$_2$ are also observed to be important constituents of ice mantles: the contributions relative to the abundance of water ice is at the level of 10-40\%  for CO (Chiar et al.\ 1995; Gibb et al.\ 2004; Chiar et al.\ 2011; \"{O}berg et al.\ 2011) and 20-30\% for CO$_2$ (Whittet et al.\ 2009; Gibb et al.\ 2004; \"{O}berg et al.\ 2011). Our models are in general agreement with these observations. There are, however, a few exceptions. The CO ice abundance in the 10 Myr delay time non-magnetic model shows a great sensitivity to temperature, with the 15 K model abundance being much lower (more than 3 orders of magnitude) than the rest of the models and observations. Similarly, the CO$_2$ abundance is decreased by much compared to observations for the 7 K versions of the 10 Myr delay and 1 Myr delay non-magnetic models, and all of the magnetic models. 
In observed clouds CH$_3$OH is also found at the few \% level relative to the water ice abundance (Boogert et al.\ 2011; \"{O}berg et al.\ 2011); in our models the CH$_3$OH abundance is at the order of $\lesssim 1\%$.

Nitrogen-bearing ices are dominated in our models by NH$_3$ ($\lesssim 10\%$ of the water ice abundance), whereas CN ice is very underabundant, consistent with observations (Gibb et al.\ 2004, Chiar et al.\ 2011). NH$_3$ ice is shown to be a good chronometer,  with its abundance almost monotonically increasing with the evolutionary timescale. HCN is systematically more abundant than HNC. N$_2$ ice is very sensitive to temperature as discussed above, and its abundance decreases with increasing temperature. 

Molecular oxygen ice abundance has a  sensitive dependence on the elemental C/O ratio (cyan and yellow bands for magnetic and non-magnetic models respectively), and its abundance decreases as oxygen elemental abundance decreases, as expected. It is this dependence that dominates the range in possible values for the O$_2$ ice abundance.

The evolutionary trends with central density followed in our models by  various mantle ice abundances are generally similar: the abundance builds up fast as a result of efficient molecule freeze-out onto grains and high enough (A$_{\rm V} > 3$) extinction, and saturates at higher values of the central density.

\subsection{Radial Abundance Profiles}
\begin{figure*}
%\plotone{plotradabund_big_final.eps}
\plotone{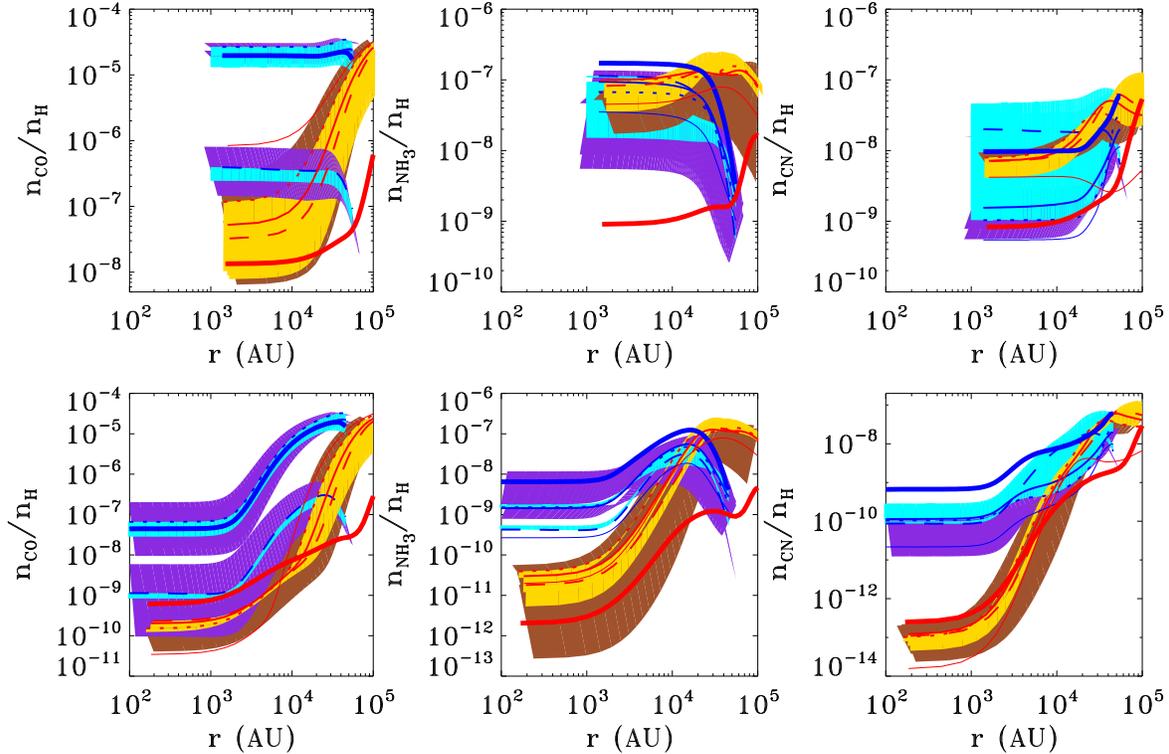}
\caption{\label{comp_rad_abund} 
Radial profiles of gas-phase abundances of three commonly observed molecules, taken when the core has reached a central density of $10^4 {\rm \, cm^{-3}}$ (upper row) and $10^6 {\rm \, cm^{-3}}$ (lower row). 
Reddish hues correspond to magnetic 
models (dashed: magnetically subcritical; solid: magnetically critical; dotted: magnetically supercritical), 
and bluish hues to non-magnetic models (dashed: 10 Myr delay; solid: 1 Myr delay; dotted: no delay). 
The thin and thick solid lines correspond to cosmic 
ray ionization rate $\zeta$ a factor of four above and below the ``reference'' value. The orange/cyan shaded areas show the effect of a varying C/O ratio for magnetic/non-magnetic runs, and correspond to a range of C/O values between $0.4$ (reference value) and $1.2$. The brown/purple shaded areas show the effect of varying temperature for magnetic/non-magnetic models and correspond to a range of $T$ values between $7$ K and $15$ K (reference value is $10$ K).}
\end{figure*}

Figure \ref{comp_rad_abund} shows radial profiles of gas-phase abundances of representative 
molecules, taken when the core has a 
central density of $10^4 {\rm \, cm^{-3}}$ (upper row) and $10^6 {\rm \, cm^{-3}}$ (lower row).
Colors and lines are the same as in Fig.~\ref{gmices}. 

At high densities (lower row), both magnetic and non-magnetic models have experienced significant central depletion in all molecules (although generally to different extent). This is consistent with the qualitative picture of Fig.~\ref{comp_cen_abund}: at high densities the molecular abundances in all models have decreased significantly. 
At lower densities however (upper row), magnetic models exhibit a qualitatively different behavior from non-magnetic models, at least for certain molecules. The difference is most striking in the case of CO, which for non-magnetic models, is centrally-peaked at a central density of $10^4 {\rm \, cm^{-3}}$. In contrast, at the same density, magnetic models have already suffered significant depletion at the center.  In the case of  NH$_3$ and CN, however, the radial profiles are quite similar (centrally peaked or flat) in both magnetic and non-magnetic models, even at low densities. As we will see in \S \ref{DM}, this effect can be used to discriminate between non-magnetic and magnetic models.

An important point to note is that the plots in Fig.~\ref{comp_rad_abund} are in a logarithmic scale, and, as a result, most of the core  mass is located in the larger scales depicted in the plots. For this reason, even in the cases where some or all models have undergone significant central depletion, when a core is not resolved the observed abundances sample mostly the outer layers where little if any depletion has occurred. This results in degeneracies among models with otherwise significant differences in central depletion, when a core is unresolved. This point is prominently demonstrated in \S \ref{totab}.

%%%
%%%
%%%
%%%

\begin{figure*}
\epsscale{0.8}
\plotone{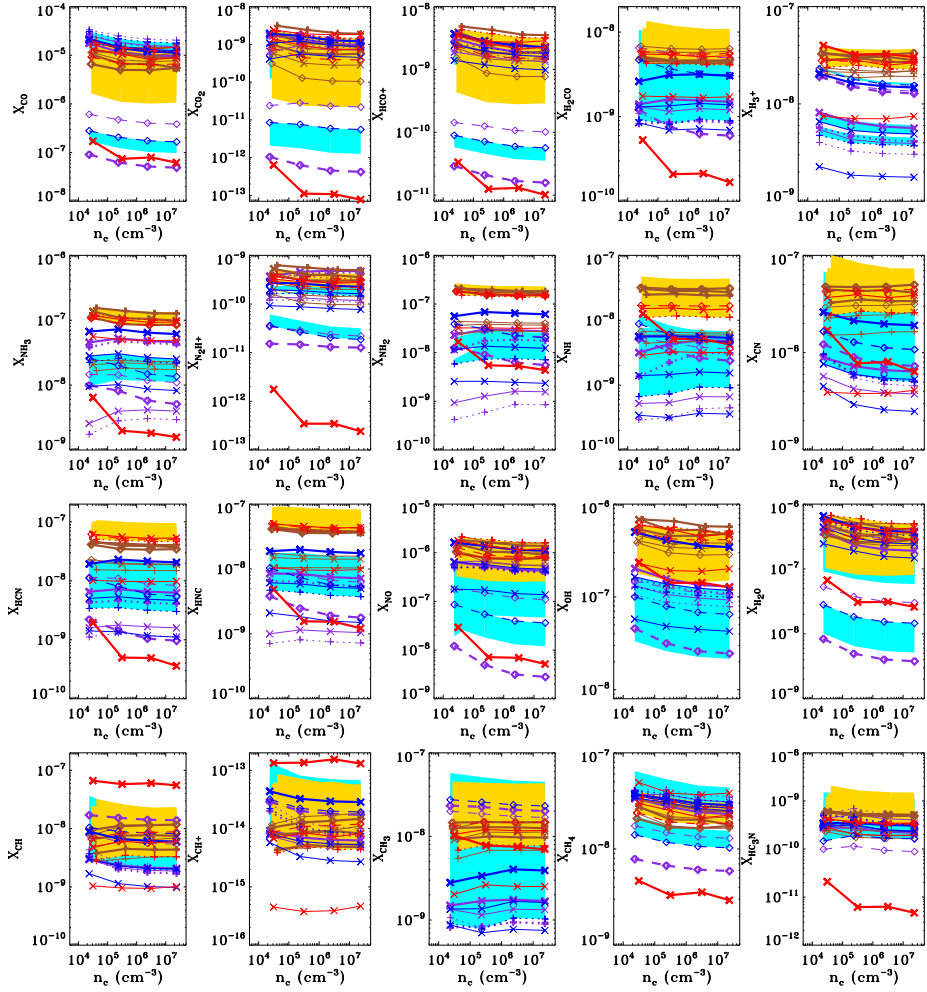}
\caption{\label{comp_tot_abund} 
Abundances {\em mass-averaged over the entire core} for representative molecules 
at four different stages in the core's evolution (quantified by the core central density), 
for different dynamical models, different C/O ratios, and different cosmic ray ionization rates. 
Reddish hues correspond to magnetic 
models (dashed: magnetically subcritical; solid: magnetically critical; dotted: magnetically supercritical), 
and bluish hues to non-magnetic models (dashed: 10 Myr delay; solid: 1 Myr delay; dotted: no delay). 
The thin and thick solid lines correspond to cosmic 
ray ionization rate $\zeta$ a factor of four above and below the ``reference'' value. The orange/cyan shaded areas show the effect of a varying C/O ratio for magnetic/non-magnetic models, and correspond to a range of C/O values between $0.4$ (reference value) and $1.2$. 
The brown/purple {\em lines}  show the effect of varying temperature for magnetic/non-magnetic models. Thin brown/purple lines:   $T = 7$ K; thick brown/purple lines:  $T = 15$ K (reference value is $10$ K).
}
\end{figure*}

\begin{figure*}
\plotone{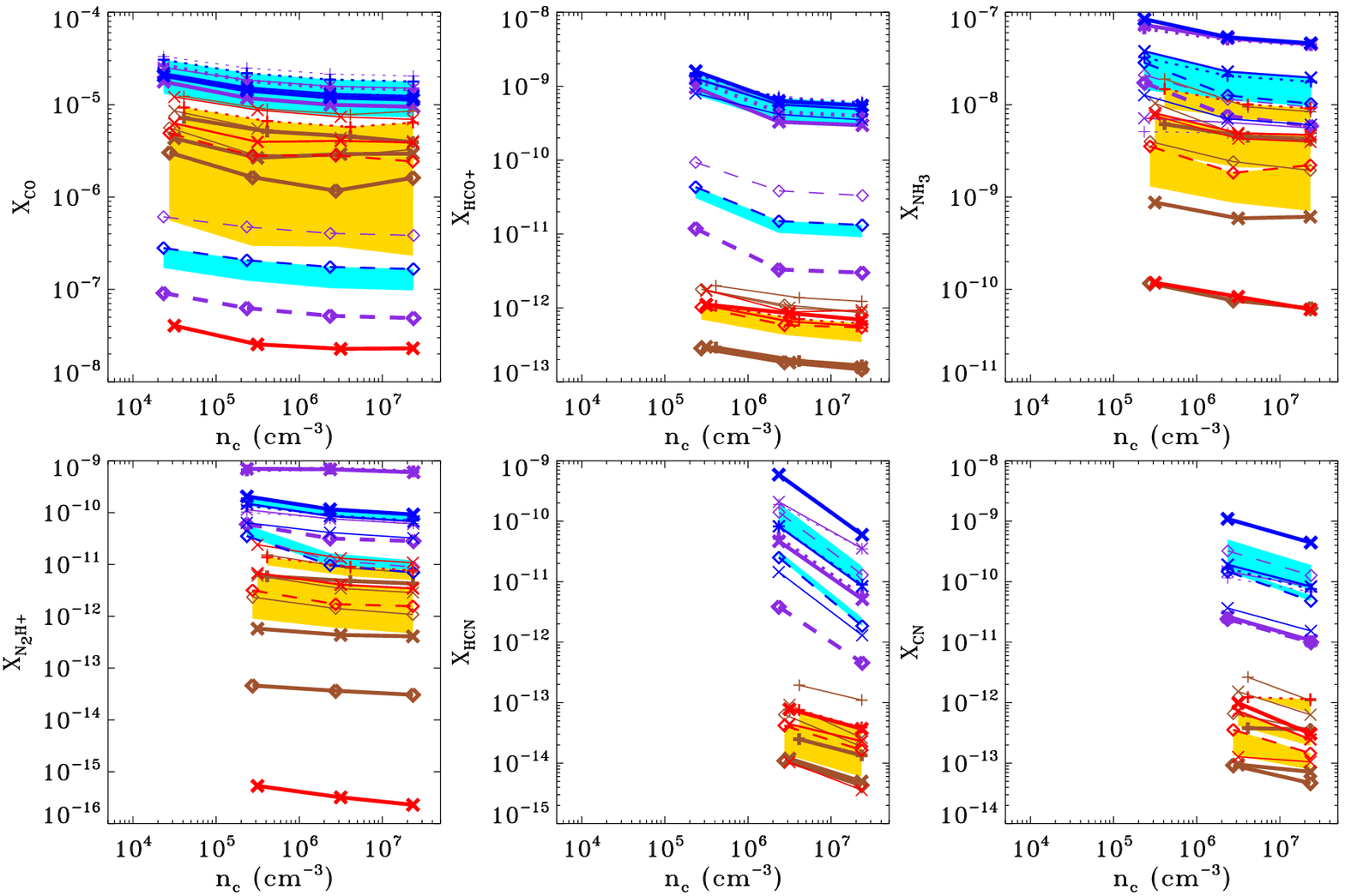}
\caption{\label{ncritt_abund} 
 As in Fig.\ \ref{comp_tot_abund}, but mass-averaged only over regions
of the core with densities above the critical density $n_{\rm crit}$ for excitation
of commonly observed lines for each molecule: CO(J=1$\rightarrow$0),
115 GHz, $n\ge n_{\rm crit} =2.18\times 10^3 {\rm \, cm^{-3}}$; 
HCO$^+$(J=1$\rightarrow$0) 89.2 GHz,
$n \ge n_{\rm crit} =1.63\times 10^5 {\rm \, cm^{-3}}$; 
NH$_3$(J,K=1,1), 23 GHz,
$n \ge 10^5 {\rm \, cm^{-3}}$, see text;
N$_2$H$^+$(J=1$\rightarrow$0), 93.17 GHz,
$n \ge n_{\rm crit} =2.25\times 10^5 {\rm \, cm^{-3}}$;
HCN(JF=12$\rightarrow$01), 88.631 GHz, 
$n \ge n_{\rm crit} =2.18\times 10^6 {\rm \, cm^{-3}}$;
CN(N=1$\rightarrow$0), 113.17 GHz, 
$n \ge n_{\rm crit} =1.3\times 10^6 {\rm \, cm^{-3}}$.
}
\end{figure*}

%%%
%%%
%%%
%%%

%%%%%%%%%%%%%%%%%%%%%%%%%%%%%%%%%%%%%%%%%%%%%%%%%%%%%%%%%%%%%%%%%%%%%%%%

\subsection{Total Abundances -- Unresolved Cores}\label{totab}

Figure \ref{comp_tot_abund} 
shows the total gas-phase 
abundances  of selected molecules in the entire core (i.e. abundances mass-averaged over the entire extent 
of the ``unresolved'' core), as a function of evolutionary stage, quantified by the core central density (x-axis). 
More advanced evolutionary stages are to the right side of the horizontal axis. The effect of C/O ratio is shown with the shaded orange and cyan regions (for magnetic and non-magnetic models respectively), while the effect of temperature is shown with thin/thick brown and purple lines (thin for $T=7$ K, thick for $T=15$ K, brown for magnetic models, purple for non-magnetic models). The effect of $\zeta$ is only shown for the reference magnetic and non-magnetic models, and is represented by the thickness of the solid blue and red lines (thicker line corresponds to higher $\zeta$).

A striking feature of the total gas-phase abundances is that  there is little evolution with central density. 
The cause of this effect is the dominant 
contribution of the outer layers of the core (the lower-density regions) to the total core mass and their 
weight in determining the total core abundances as would be observed in unresolved cores. Since the 
abundances in the outer layers of the cores do not evolve much, the total core abundances also remain 
relatively stable with evolutionary stage. 

Because of the power-law abundance profiles, 
this feature is generally preserved even if one averages only over radii with
densities above the critical density for excitation of a specific
molecular line instead of over the entire core. 
Of course, the fact that a specific line is observed already places a
lower limit on the central density of the core, equal to $n_{\rm
  crit}$ for the specific line. However, little evolution of the
abundance with time is seen beyond that limit, unless $n_{\rm crit}$
is very high ($\ge 10^6 {\rm \, cm^{-3}}$). 
This can be
seen in Fig.\ \ref{ncritt_abund} , where we have plotted abundances
of selected molecules averaged over regions with densities higher than
the critical density $n_{\rm crit}$ for the following molecular lines: 
CO(J=1$\rightarrow$0),
115 GHz, $n\ge n_{\rm crit} =2.18\times 10^3 {\rm \, cm^{-3}}$; 
HCO$^+$(J=1$\rightarrow$0) 89.2 GHz,
$n \ge n_{\rm crit} =1.63\times 10^5 {\rm \, cm^{-3}}$; 
NH$_3$(JK=11), 23 GHz,
$n \ge 10^5 {\rm \, cm^{-3}}$[Although the critical density
  for this transition is much lower ($\sim 2\times 10^3 {\rm \,
    cm^{-3}}$), this line requires much higher densities before the
  line is thermalized (Evans 1989)];
N$_2$H$^+$(J=1$\rightarrow$0), 93.17 GHz,
$n \ge n_{\rm crit} =2.25\times 10^5 {\rm \, cm^{-3}}$;
HCN(JF=12$\rightarrow$01), 88.631 GHz, 
$n \ge n_{\rm crit} =2.18\times 10^6 {\rm \, cm^{-3}}$;
and CN(N=1$\rightarrow$0), 113.17 GHz, 
$n \ge n_{\rm crit} =1.3\times 10^6 {\rm \, cm^{-3}}$.
These critical densities are for $T =10$ K. All molecular data are taken
from the Leiden Atomic and Molecular Database (Sch\"{o}ier et al.
2005).\footnote{http://www.strw.leidenuniv.nl/$\sim$moldata/}

The large range of possible abundance values for different models 
at a fixed evolutionary stage (central density) demonstrates the large potential of chemical differentiation in prestellar cores. Even cores that appear identical in, e.g., their total column density profiles, can exhibit large variations in their molecular abundances as a result of a diversity in other local physical parameters (temperature, ionization rate, initial elemental abundance ratios, and dynamics). 

However, the large variation in molecular abundances among different models is neither monotonic nor systematic, and the abundances resulting from different parameters are not well-separated. Instead, we often see a nonlinear dependence on certain parameters (a characteristic example is the ionization rate), as well as largely overlapping ranges, resulting in strong model degeneracies. We conclude that discrimination between models is not straight-forward, but instead requires the use of a variety of different observations and constraints, in combination with a statistical study accounting for as many different molecular abundances as possible. In the next sections, we will discuss certain quantities, such as depletion measures and abundance ratios, which also hold promise as tools for the discrimination among dynamical models and physical parameters. 

We caution the reader that the abundances in Fig.~\ref{comp_tot_abund} are not to be compared directly with abundances derived from observations beyond an uncertainty of a factor of at least several. There are various reasons for possible discrepancies. First of all, the mass-averaged abundances presented here are not appropriate when the core is at least partly resolved; in this case, a more detailed study is needed, starting with the radial profile at a specific evolutionary snapshot for each model, and properly accounting for the beam size and the physical scale it corresponds to at the distance of the core. Second, chemical reaction rates have appreciable uncertainties (Wakelam et al.~2010). We have not studied the effect of such uncertainties here, as our purpose was to investigate the effect of varying the core dynamics and other physical parameters  on molecular abundances. Finally, obtaining a molecular abundance from observations involves several approximations as well as assuming a  value for the H$_2$ column density (which is not always well-constrained). Nevertheless, we have verified, for the following cores, that molecular abundances obtained from observations are generally within our model ranges, or only deviate from it within uncertainties: L1498 (Tafalla et al.~ 2004, 2006); L1517B (Tafalla et al.~2004, 2006; Crapsi 2005); L1544 (Caselli et al.~1999; Jorgensen et al.~2004); L134N (Wakelam et al.~2006); TMC-1(CP) (Smith, Herbst \& Chang 2004); L1689B (Jorgensen et al.~2004). 
For an improved comparison with observations, the appropriate methodology would be to start from our dynamical/chemical models, perform radiative transfer calculations, and directly compare predicted line intensities with observed ones. We plan to return to such a comparison in a future study.

\subsection{Depletion Measures}\label{DM}

\begin{figure*}
\epsscale{0.8}
\plotone{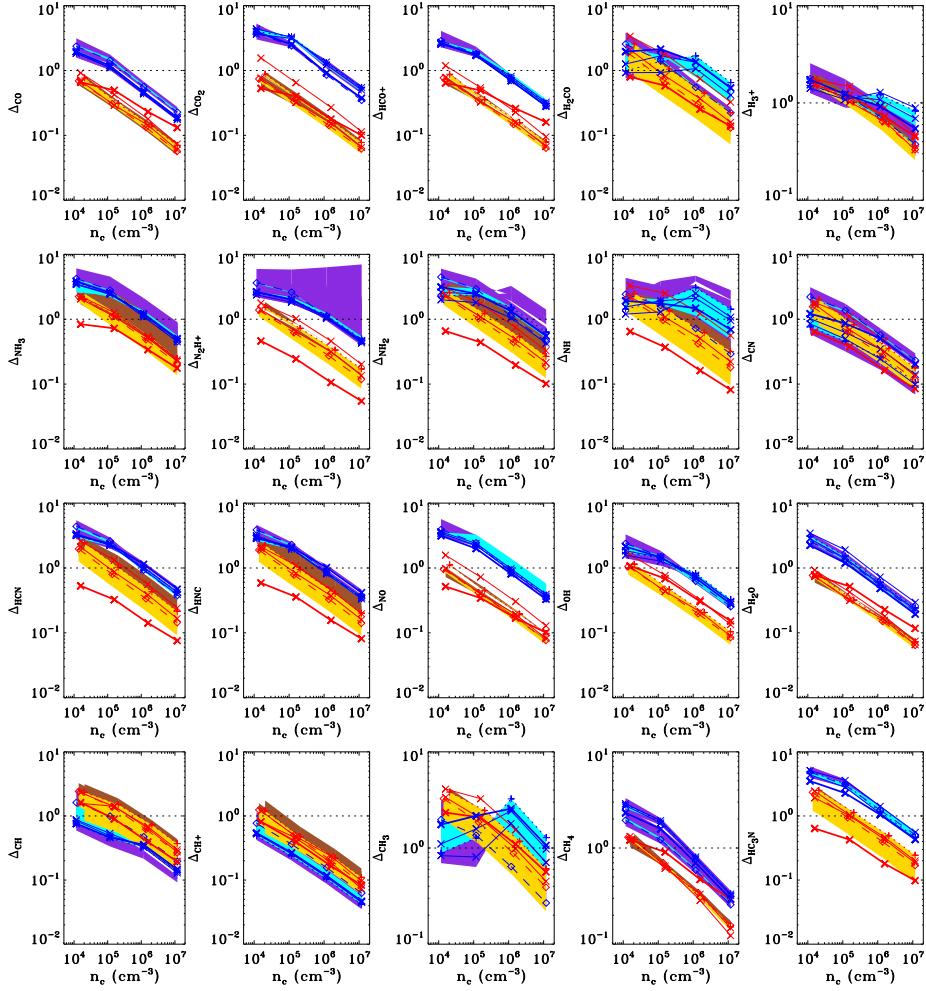}
\caption{\label{delta} Central deviation of molecular abundance with respect to the average abundance over the entire core ($\Delta$, see text), as a function of central density. The dotted line corresponds to $\Delta = 1$ (no central enhancement or depletion). 
Reddish hues correspond to magnetic 
models (dashed: magnetically subcritical; solid: magnetically critical; dotted: magnetically supercritical), 
and bluish hues to non-magnetic models (dashed: 10 Myr delay; solid: 1 Myr delay; dotted: no delay). 
The thin and thick solid lines correspond to cosmic 
ray ionization rate $\zeta$ a factor of four above and below the ``reference'' value. The shaded areas show the effect of a varying C/O ratio, and correspond to a range of C/O values between $0.4$ (reference value) and $1.2$.
The brown/purple shaded areas show the effect of varying temperature for magnetic/non-magnetic models and correspond to a range of $T$ values between $7$ K and $15$ K (reference value is $10$ K).}
\end{figure*}

The qualitatively different behavior between the radial profiles of magnetic and non-magnetic models seen in Fig.~\ref{comp_rad_abund} can be used to discriminate between models. We define the {\em central deviation} $\Delta$ (central depletion or enhancement) of the molecular abundance as the ratio between the integrated abundance through the central $30\%$\footnote{We have confirmed that the results are similar if this fraction is changed to 10\% or 20\%.} of the radial extent of the core over the integrated abundance over the entire core. 

Figure \ref{delta} shows the dependence of $\Delta$  on evolutionary stage (here quantified by the value of the central density) for all models (colors and lines as in Fig.~\ref{comp_rad_abund}). The horizontal dotted line marks the value $\Delta =1$. Values of $\Delta$ above the line indicate that the particular molecule is centrally enhanced. Values of $\Delta$ close to the line indicate a flat abundance profile, while values of $\Delta$ below the line correspond to centrally depleted molecules.

Despite the significant overlap between magnetic and non-magnetic models both in the radial profiles and the total abundances discussed in Figs.~\ref{comp_rad_abund} and \ref{comp_tot_abund}  respectively, for many molecules the magnetic and non-magnetic models not only are well separated, but also show little sensitivity to model parameters such as the evolutionary timescale, the temperature, the C/O ratio, and the cosmic-ray ionization rate. 
Molecules which show such good separation between magnetic and non-magnetic models include CO, CO$_2$, HCO$^+$, NO, OH, H$_2$), and HC$_3$N. In contrast, H$_3^+$ and CN show a particularly degenerate behavior between different models. 

\begin{figure}
\plotone{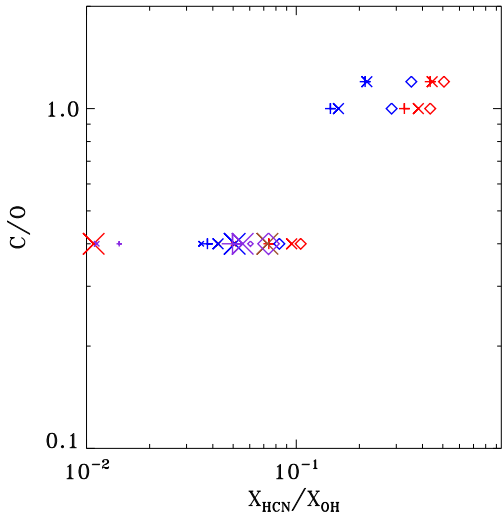}
\caption{\label{ratioC2O} 
Gas-phase abundance ratio between HCN and OH, mass-averaged over the entire core, which has good potential for discriminating between C/O ratio values. Red/blue symbols correspond to magnetic/non-magnetic models with $T=10$ K.  Diamonds': ``slow'' models; X: ``reference'' models; crosses: ``fast'' models. The size of the red/blue symbols represents the value of $\zeta$ (larger symbols correspond to larger $\zeta$).  Brown/purple symbols correspond to magnetic/non-magnetic models with $T=7$ K (small symbols) and $T=15$ K (large symbols), and a ``reference'' value for $\zeta$. The C/O ratio of each model is shown on the vertical axis, while the value of the abundance ratio is shown on the horizontal axis. Points correspond to a central $H_2$ density of $10^6 {\rm cm^{-3}}$, however the core-average abundance ratios evolve little with density.}
\end{figure} 

\begin{figure*}
\plotone{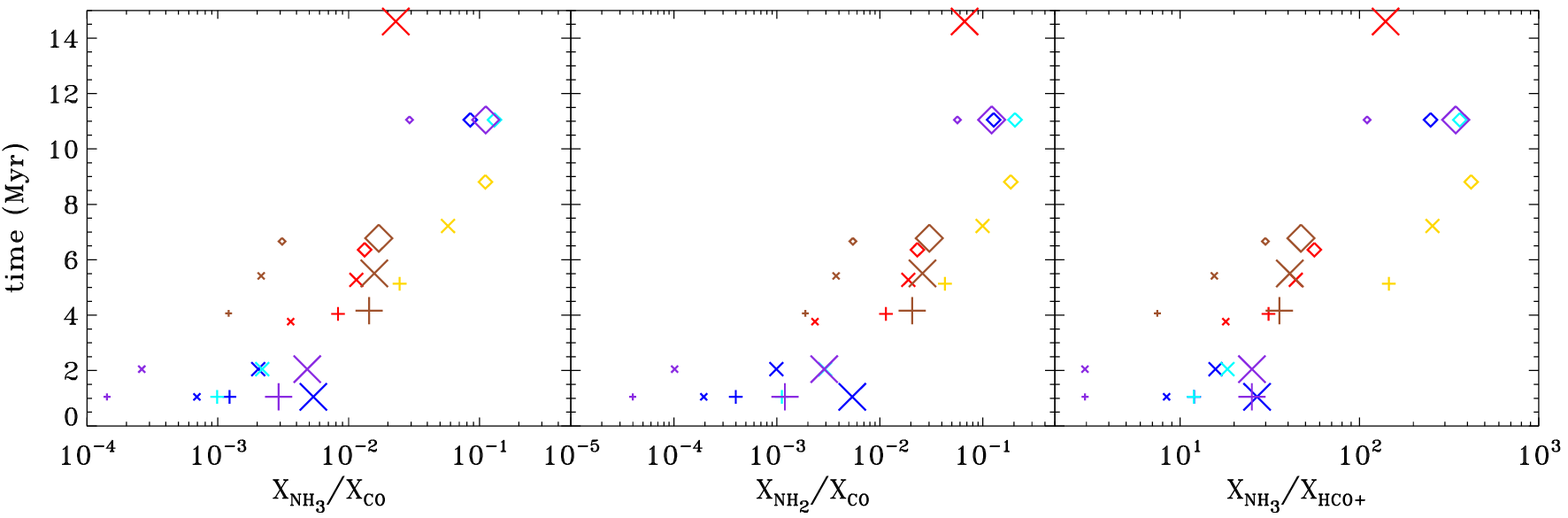}
\caption{\label{ratioTime} 
Gas-phase abundance ratios, mass-averaged over the entire core, with high potential for discrimination among evolutionary timescales for the core. Diamonds': ``slow'' models; X: ``reference'' models; crosses: ``fast'' models.  Red/blue symbols correspond to magnetic/non-magnetic models with $T=10$ K and C/O =0.4. The size of the red/blue symbols represents the value of $\zeta$ (larger symbols correspond to larger $\zeta$). Orange/cyan points correspond to magnetic/non-magnetic models with $T=10$ K, C/O=1.2, and a ``reference'' value for $\zeta$. 
 Brown/purple symbols correspond to magnetic/non-magnetic models with $T=7$ K (small symbols) and $T=15$ K (large symbols), C/O=0.4, and a ``reference'' value for $\zeta$.
 The time it takes the core to reach a central density of $10^6 {\rm cm^{-3}}$ is shown on the vertical axis, while the value of the abundance ratio is shown on the horizontal axis. Points correspond to a central $H_2$ density of $10^6 {\rm cm^{-3}}$, however the core-average abundance ratios evolve little with density.}
\end{figure*}

For non-magnetic models, CO-bearing molecules, and in particular CO, CO$_2$ and HCO$^+$, tend to show no depletion ($\Delta>1$) at early stages of the collapse, and only become depleted when the central density reaches $10^5$ - $10^6 {\rm \, cm^{-3}}$. In contrast, for magnetic models, the same molecules are already centrally depleted by central densities of $\sim 10^4  {\rm \, cm^{-3}}$. 

 NH$_3$ on the other hand shows no sign of central depletion at central densities of $\sim 10^4  {\rm \, cm^{-3}}$ for both magnetic and non-magnetic models. It only becomes depleted at later stages of the collapse ($10^6$ - $10^7 {\rm \, cm^{-3}}$ for non-magnetic models, and  $10^5$ - $10^6 {\rm \, cm^{-3}}$ for magnetic models). N$_2$H$^+$ shows a similar behavior, however its behavior is complicated by its sensitivity to temperature in the case of non-magnetic models (see also discussion in \S \ref{evca}). Note also that in all N-bearing molecules, the high cosmic ray ionization rate magnetic model is always depleted, as a result of the very long evolutionary timescale (see also discussion of Fig.\ref{comp_cen_abund}).  This difference in behavior between CO and   NH$_3$ is consistent with observations (e.g., Caselli et al.~1999; Bacmann et al.~2003; Crapsi et al.~2007).

\subsection{Abundance ratios of molecules with discriminatory potential}

Ratios between abundances of different molecules have been long considered as promising tools to overcome model uncertainties and degeneracies. However, most abundance ratios exhibit large variations and no consistent trend with any single model parameter when other parameters are varied. We have identified several gas-phase abundance ratios (averaged over the entire core) that are most affected by a single model parameter and show little variation as a result of the rest and thus show promising discriminatory potential. We plot these  in Figs.~\ref{ratioC2O} and \ref{ratioTime}.

Figure \ref{ratioC2O} shows the values of the abundance ratio between HCN and OH, mass-averaged over the entire core, for our various models. This ratio can serve as an indicator of the C/O ratio. Different models are plotted with different symbols as discussed in the caption. The C/O ratio, shown in the vertical axis, exhibits a trend (increases with increasing value of each abundance ratio). Similarly, Figure \ref{ratioTime} shows three abundance ratios that trace the evolutionary timescale of a core. These ratios are  ${\rm X_{NH_3}/X_{CO}}$, ${\rm X_{NH_2}/X_{CO}}$, and ${\rm X_{NH_3}/X_{HCO^+}}$. All the ratio values are taken at the time when the central density in each model has reached $10^6 {\rm \, cm^{-3}}$. However, the averaged abundance over the entire core shows little time evolution, as shown in Fig.~\ref{comp_tot_abund}.

 Each abundance ratio examined in these two figures shows an appreciable scatter at a fixed value of the parameter that it can help constrain (C/O ratio or evolutionary timescale, respectively). However, a general trend is obvious, and there are significant parts of the C/O or timescale parameter space that can be excluded with measurements of each ratio.

\section{Discussion and Conclusions}\label{thedisc}
 
We have combined dynamical and non-equilibrium chemical modeling of both magnetic and non-magnetic 
contracting molecular cloud cores, in order to predict molecular abundances as a function of core 
evolutionary stage, and identify observables with high promise in discriminating among core models. 

To this end, we have undertaken an extensive parameter study, following a total of 34 different models, 
and varying: \\
(a) the core evolutionary timescale, controlled by the {\em collapse delay time}, in the case of 
non-magnetic models, and the initial {\em mass-to-magnetic-flux ratio}, i.e. the amount of magnetic 
support against gravity, in the case of magnetic models; \\
(b) the elemental C/O ratio in the core;  \\
(c) the cosmic-ray ionization rate, which affects the chemistry in both magnetic and non-magnetic models, 
but also the evolutionary timescale in magnetic models, by affecting the extent of coupling between the 
magnetic field and the neutral fluid; and \\
(d) the temperature. 

Our main results can be summarized as follows. 
We have explicitly demonstrated that many reactions in our chemical network are out of 
equilibrium during the dynamical evolution of the core, necessitating the inclusion of non-equilibrium 
chemistry and the coupled modeling of chemistry and dynamics to properly follow the molecular abundances. 

For different dynamical models, the molecular abundances at the center of the core are not the same, 
even for identical values of the central number density; rather, the amount of time that a model core 
has spent {\em at each density} is what determines the actual molecular abundances 
at each evolutionary stage. Therefore it is not possible to properly account for magnetic effects in core 
chemistry by using a collapse retardation factor in non-magnetic models. Even if 
the overall final age of the core is the same, the fact that the non-magnetic core model has spent 
very different fractions of this age at various values of the central density than a magnetic core model
has, results in significantly different molecular abundances. 

Abundances in magnetic models peak at much lower densities and decrease much faster at higher densities. 
The evolution of grain mantle ice abundances is qualitatively similar in magnetic and non-magnetic models, 
with a fast initial increase due to freeze-out, and an abundance saturation at higher densities and later times. 
The value of that saturation however is different in different models. 
We find that the abundance of NH$_3$ ice scales roughly with the evolutionary timescale of the model, varying by a factor of 4 for our model range, which corresponds to an evolutionary timescale range of about 10 Myr. NH$_3$ ice is thus a potentially  important chronometer. 

Radial profiles of gas-phase abundances reveal that the most significant difference between magnetic and 
non-magnetic models is at the central parts of the core, while most of the core extent tends to 
evolve much less and is much more similar among different models. However, as only a small fraction of the mass is concentrated at the central layers of the core, the central differences are smeared out in 
unresolved  cores. As a result, the evolution of the total, mass-averaged molecular abundances with 
central density in any particular model is much milder, if at all present. The situation is similar 
with the total abundances of grain mantle ices. 

Differences in the C/O ratio tend to affect mostly the abundance of the oxygen-bearing molecules, while the 
cosmic-ray ionization rate $\zeta$ has its largest effect on the nitrogen-bearing molecules. In the case 
of magnetic models, the direct effect of the cosmic-ray ionization rate on chemistry is compounded with 
its indirect effect on core evolution timescale, and as a result magnetic models are much more sensitive 
to changes in $\zeta$.

The many, sometimes competing, sometimes compounded, effects of different model parameters on molecular 
abundances result in severe degeneracies among different models. For this reason, we have identified 
{\em molecular abundance ratios} which may help to constrain, even in unresolved cores, specific 
model parameters: the evolutionary timescale (ratios between NH$_3$ and CO; NH$_2$ and CO; NH$_3$ and HCO$^+$) 
and the C/O ratio (ratio between HCN and OH).

In addition, we have demonstrated that the {\em central deviation} $\Delta$ (central depletion or enhancement) of molecules can be another useful tool in model discrimination. While for many molecules values of $\Delta$ are degenerate among different models, there are several molecules for which there is a clear separation between magnetic and non-magnetic models. For these molecules the value of $\Delta$ even shows very little sensitivity to other model parameters. Measurements of the central depletion of such molecules therefore have an excellent potential for constraining dynamical models of core evolution. For this reason, it is very important to place adequate emphasis into understanding differential depletion and grain chemistry effects as much as possible, and thus gain confidence in the chemistry part of these calculations, which still features significant uncertainties (eg., Wakelam et al.\ 2010). 

The dynamical models we have used in this study have low dimensionality: the non-magnetic models are 
spherically symmetric (one-dimensional), while the magnetic ones are axially symmetric, and integrated 
over the z-direction (1.5 dimensions).  We note that as a result of the
distinct geometries, the magnetic model does not converge to the
non-magnetic model in the limit of zero magnetic field. A zero
magnetic field disk model cannot be followed with the current
numerical scheme, as the thin-disk approximation breaks down in the
absence of a magnetic field. 
Although higher-dimensional calculations are desirable 
(e.g., van Weeren et al.~2009), it is currently not possible to perform a parameter study as extensive 
as the one undertaken here using higher-dimensional models: a single model in this study 
(including non-equilibrium chemistry) requires more than 800 CPU hours for the magnetic runs and more 
than 200 CPU hours for the non-magnetic runs. We are however in the
process of implementing true 2D calculations for select cases
presented here, to assess the effect of low dimensionality on our calculations.

Because the model cores are not multi-dimensional, multiplicity or fragmentation is by definition neither 
accounted for nor allowed. This however is not a severe disadvantage in our case, as multiplicity at 
these early stages of collapse and for the low densities considered here (up to $10^7 { \rm \, cm^{-3}}$) 
is not observed (Schnee et al.~2010). 

Finally our work has not accounted for turbulent mixing; the flows in
our model are laminar. Any turbulent mixing would result in making the
radial profiles shallower (Xie et al. 1995).
However, dense molecular cloud cores are generally not observed to have significant amounts of turbulence since 
their line widths are approximately sonic (Myers 1983; Barranco \& Goodman 1998; 
Goodman et al.\ 1998;  Kirk, Johnstone, \& Tafalla 2007). 

\acknowledgements{We thank Paul Goldsmith, Talayeh Hezareh, and the anonymous referee
  for insightful comments that have improved this paper. This work was carried out at the 
Jet Propulsion Laboratory, California Institute of Technology, under a contract with the National Aeronautics 
and Space Administration.  \copyright 2012. All rights reserved.}

%%%%%%%%%% Appendix
\appendix

\section{Code Tests}
\label{codetests}

As a test of the dynamical part of our code we have checked that, for the (more complex) magnetic model, 
our code returns the same results as Basu \& Mouschovias (1994) when identical initial conditions are used. 

\begin{figure}[h]
\begin{center}
\resizebox{2in}{!}{
\plotone{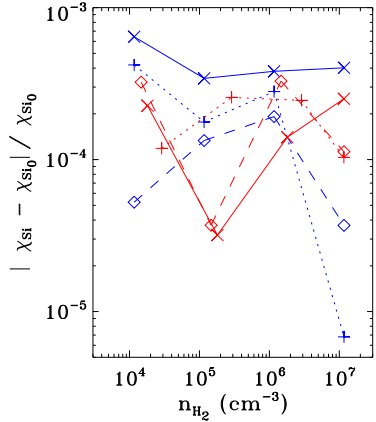}}
\caption{\label{test_element} 
Fractional change of total elemental abundance of Si ($\chi_{Si}$) with respect to the initial one ($\chi_{Si_0}$)
versus central density in collapsing core models. Pure hydrodynamical model cores are shown with blue symbols
and lines, while MHD core models are shown with red symbols and lines as in Fig.~\ref{n_vs_t}. 
}
\end{center}
\end{figure}

To confirm that the chemistry implementation in our code behaves as
expected, we have checked that the elemental abundances are conserved
in our simulations. Figure \ref{test_element} shows the fractional 
changes in the abundance of the most underabundant element Si 
(abundance of $10^{-8}$, see Table \ref{table2}) as a fraction of 
central volume density, for magnetic (red lines) and non-magnetic (blue lines) of different mass-to-magnetic-flux ratios and initial delays, respectively. The fractional change always remains $< 10^{-3}$ for all 
simulations. We thus confirm that  the codes conserve elemental abundances better than one part in a 
thousand of the total abundance even for the most underabundant elements. 

\begin{figure}[b]
\begin{center}
\resizebox{5in}{!}{
\plottwo{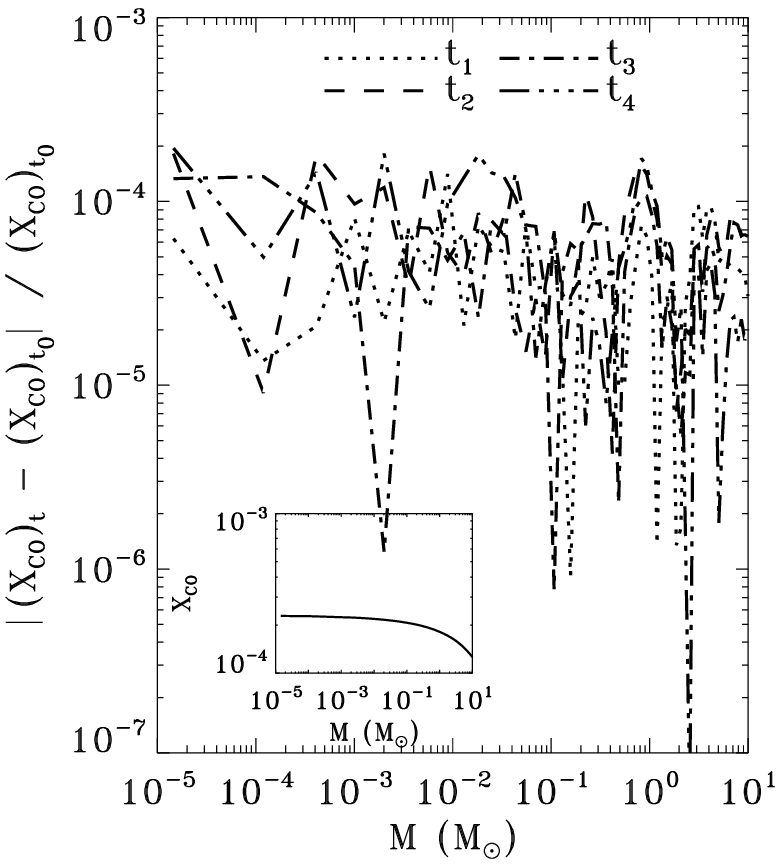}{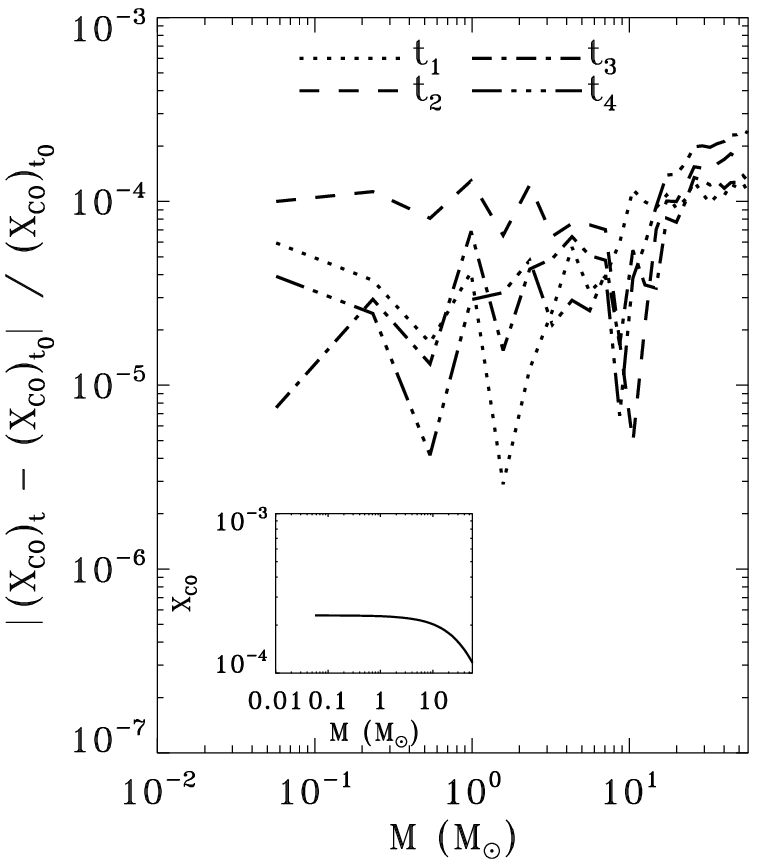}
} \caption{\label{test_CO} 
Fractional change of the CO abundance, as a function of mass, at different times of the
collapse, normalized to the initial time $t_0$. 
Left panel: non-magnetic reference model; Right panel: magnetic
reference model. The inset shows the initial abundance profile. 
}
\end{center}
\end{figure}

In addition, we have performed a test on the advection of chemical species. 
Retaining identical initial and boundary conditions as in the main
runs, in both magnetic and non-magnetic models, we allowed the
chemistry 
to evolve for one million years before any dynamical evolution 
was allowed to take place. In this way, an abundance profile for the various molecules was developed. During 
the next step of the test, the dynamics were turned on, while the chemical reactions were turned off, so that 
molecules were advected, but their abundances could not be altered due to sources or sinks.  Advection should not 
be changing the distribution of abundance with mass, and the abundances within a single mass element should 
be conserved. The only expected change, as the contraction of the core proceeds, is  the distribution of mass 
elements with scale. The results are presented in  Fig.~\ref{test_CO}. 
The left panel corresponds to the non-magnetic reference model and the right panel to the magnetic reference 
model. The plots show the relative
error (normalized to the initial time) of the abundance of CO for
different mass elements for different snapshots in time after the
dynamics are turned on. The CO abundance profile is shown in the
insets.  We conclude that  
no spurious numerical alteration in molecule abundances is induced because of advection, as expected for 
a fully conservative advection scheme such as the one implemented here (Morton, Mouschovias, \& Ciolek  1994).


\begin{thebibliography}{1}
\bibitem[Aikawa et al.(2001)]{2001ApJ...552..639A} Aikawa, Y., Ohashi, N., 
Inutsuka, S.-i., Herbst, E., \& Takakuwa, S.\ 2001, \apj, 552, 639. %Molecular Evolution in Collapsing Prestellar Cores.

\bibitem[Aikawa et al.(2003)]{2003ApJ...593..906A} Aikawa, Y., Ohashi, N., 
\& Herbst, E.\ 2003, \apj, 593, 906. %Molecular Evolution in Collapsing Prestellar Cores. II. The Effect of Grain-Surface Reactions. 

\bibitem[Aikawa et al.(2005)]{2005ApJ...620..330A} Aikawa, Y., Herbst, E., 
Roberts, H., \& Caselli, P.\ 2005, \apj, 620, 330. 
%Molecular Evolution in Collapsing Prestellar Cores. III. Contraction of a Bonnor-Ebert Sphere.

\bibitem[Basu 
\& Mouschovias(1995)]{1995ApJ...452..386B} Basu, S., \& Mouschovias, T.~C.\ 1995, \apj, 452, 386 

\bibitem[Bacmann et 
al.(2000)]{2000A&A...361..555B} Bacmann, A., Andr{\'e}, P., Puget, J.-L., Abergel, A., Bontemps, S., \& Ward-Thompson, D.\ 2000, \aap, 361, 555. 
%An ISOCAM absorption survey of the structure of pre-stellar cloud cores. 

\bibitem[Bacmann et al.(2003)]{2003ApJ...585L..55B} Bacmann, A., Lefloch, 
B., Ceccarelli, C., et al.\ 2003, \apjl, 585, L55 

\bibitem[Barranco 
\& Goodman(1998)]{1998ApJ...504..207B} Barranco, J.~A., \& Goodman, A.~A.\ 1998, \apj, 504, 207. 
%Coherent Dense Cores. I. NH 3 Observations.

\bibitem[Basu \& Mouschovias (1994)]{BM94}
Basu, S. \& Mouschovias, T. Ch.\ 1994, ApJ, 432, 720

\bibitem[Bergin 
\& Langer(1997)]{1997ApJ...486..316B} Bergin, E.~A., \& Langer, W.~D.\ 1997, \apj, 486, 316.
%Chemical Evolution in Preprotostellar and Protostellar Cores.

\bibitem[Bergin 
\& Tafalla(2007)]{2007ARA&A..45..339B} Bergin, E.~A., \& Tafalla, M.\ 2007, \araa, 45, 339. 
%Cold Dark Clouds: The Initial Conditions for Star Formation.

\bibitem[Bohlin et al.(1978)]{1978ApJ...224..132B} Bohlin, R.~C., Savage, 
B.~D., \& Drake, J.~F.\ 1978, \apj, 224, 132 

\bibitem[Boogert et al.(2011)]{2011ApJ...729...92B} Boogert, A.~C.~A., et 
al.\ 2011, \apj, 729, 92 

\bibitem[Caselli et al.(1999)]{1999ApJ...523L.165C} Caselli, P., Walmsley, 
C.~M., Tafalla, M., Dore, L., \& Myers, P.~C.\ 1999, \apjl, 523, L165 

\bibitem[Crapsi et 
al.(2007)]{2007A&A...470..221C} Crapsi, A., Caselli, P., Walmsley, M.~C., \& Tafalla, M.\ 2007, \aap, 470, 221 

%\bibitem[Broderick et al.(2007)]{2007ApJ...671.1832B} Broderick, A.~E., 
%Keto, E., Lada, C.~J., \& Narayan, R.\ 2007, \apj, 671, 1832. 
%%Oscillating Starless Cores: Nonlinear Regime.

%\bibitem[Caselli et al.(2002)]{2002ApJ...572..238C} Caselli, P., Benson, 
%P.~J., Myers, P.~C., \& Tafalla, M.\ 2002, \apj, 572, 238.
%%Dense Cores in Dark Clouds. XIV. N2H+ (1-0) Maps of Dense Cloud Cores.

\bibitem[Chang et 
al.(2007)]{2007A&A...469..973C} Chang, Q., Cuppen, H.~M., \& Herbst, E.\ 2007, \aap, 469, 973 

\bibitem[Chiar et al.(1995)]{1995ApJ...455..234C} Chiar, J.~E., Adamson, 
A.~J., Kerr, T.~H., \& Whittet, D.~C.~B.\ 1995, \apj, 455, 234 

\bibitem[Chiar et al.(2011)]{2011ApJ...731....9C} Chiar, J.~E., et al.\ 
2011, \apj, 731, 9 

\bibitem[Ciolek 
\& Mouschovias(1993)]{1993ApJ...418..774C} Ciolek, G.~E., \& Mouschovias, T.~Ch.\ 1993, \apj, 418, 774 

\bibitem[Crapsi et al.(2005)]{2005ApJ...619..379C} Crapsi, A., Caselli, P., 
Walmsley, C.~M., Myers, P.~C., Tafalla, M., Lee, C.~W., 
\& Bourke, T.~L.\ 2005, \apj, 619, 379 

\bibitem[Desch 
\& Mouschovias(2001)]{2001ApJ...550..314D} Desch, S.~J., \& Mouschovias, T.~Ch.\ 2001, \apj, 550, 314 

\bibitem[Evans(1989)]{1989RMxAA..18...21E} Evans, N.~J., II 1989, RMxAA, 
18, 21 

\bibitem[Evans et al.(2001)]{2001ApJ...557..193E} Evans, N.~J., II, 
Rawlings, J.~M.~C., Shirley, Y.~L., \& Mundy, L.~G.\ 2001, \apj, 557, 193. 
%Tracing the Mass during Low-Mass Star Formation. II. Modeling the Submillimeter Emission from Preprotostellar Cores. 

%\bibitem[Falgarone et 
%al.(2008)]{2008A&A...487..247F} Falgarone, E., Troland, T.~H., Crutcher, R.~M., \& Paubert, G.\ 2008, \aap, 487, 247. 
%%CN Zeeman measurements in star formation regions.

\bibitem[Fiedler 
\& Mouschovias(1993)]{1993ApJ...415..680F} Fiedler, R.~A., \& Mouschovias, T.~C.\ 1993, \apj, 415, 680.
%Ambipolar Diffusion and Star Formation: Formation and Contraction of Axisymmetric Cloud Cores. II. Results.

\bibitem[Flower et al.(2005)]{Fl2005} 
Flower, D.~R., Pineau Des For{\^e}ts, G., \& Walmsley, C.~M.\ 2005, \aap, 436, 933 

\bibitem[Garrod 
\& Pauly(2011)]{2011ApJ...735...15G} Garrod, R.~T., \& Pauly, T.\ 2011, \apj, 735, 15 

\bibitem[Gibb et al.(2004)]{2004ApJS..151...35G} Gibb, E.~L., Whittet, 
D.~C.~B., Boogert, A.~C.~A., \& Tielens, A.~G.~G.~M.\ 2004, \apjs, 151, 35 

\bibitem[Gong 
\& Ostriker(2009)]{2009ApJ...699..230G} Gong, H., \& Ostriker, E.~C.\ 2009, \apj, 699, 230 

%\bibitem[Goodman et al.(1993)]{1993ApJ...406..528G} Goodman, A.~A., Benson, 
%P.~J., Fuller, G.~A., \& Myers, P.~C.\ 1993, \apj, 406, 528.
%Dense cores in dark clouds. VIII - Velocity gradients.

\bibitem[Goodman et al.(1998)]{1998ApJ...504..223G} Goodman, A.~A., 
Barranco, J.~A., Wilner, D.~J., \& Heyer, M.~H.\ 1998, \apj, 504, 223.
%Coherence in Dense Cores. II. The Transition to Coherence.

\bibitem[Hartmann et al.(2001)]{2001ApJ...562..852H} Hartmann, L., 
Ballesteros-Paredes, J., \& Bergin, E.~A.\ 2001, \apj, 562, 852.
%Rapid Formation of Molecular Clouds and Stars in the Solar Neighborhood.

\bibitem[Hasegawa 
\& Herbst(1993)]{1993MNRAS.261...83H} Hasegawa, T.~I., \& Herbst, E.\ 1993, \mnras, 261, 83. 
%New gas-grain chemical models of quiescent dense interstellar clouds - The effects of H2 tunnelling reactions and cosmic ray induced desorption.

\bibitem[Hassel et 
al.(2010)]{2010A&A...515A..66H} Hassel, G.~E., Herbst, E., \& Bergin, E.~A.\ 2010, \aap, 515, A66 

\bibitem[Heitsch 
\& Hartmann(2008)]{2008ApJ...689..290H} Heitsch, F., \& Hartmann, L.\ 2008, \apj, 689, 290. 
%Rapid Molecular Cloud and Star Formation: Mechanisms and Movies.

\bibitem[Hezareh et al.(2008)]{2008ApJ...684.1221H} Hezareh, T., Houde, M., 
McCoey, C., Vastel, C., \& Peng, R.\ 2008, \apj, 684, 1221 

%\bibitem[Hogerheijde 
%\& van der Tak(2000)]{2000A&A...362..697H} Hogerheijde, M.~R., \& van der Tak, F.~F.~S.\ 2000, \aap, 362, 697 

\bibitem[J{\o}rgensen et 
al.(2004)]{2004A&A...416..603J} J{\o}rgensen, J.~K., Sch{\"o}ier, F.~L., \& van Dishoeck, E.~F.\ 2004, \aap, 416, 603 

%\bibitem[Keto et al.(2006)]{2006ApJ...652.1366K} Keto, E., Broderick, 
%A.~E., Lada, C.~J., \& Narayan, R.\ 2006, \apj, 652, 1366. 
%Oscillations of Starless Cores.

%\bibitem[Kessel et al. 1998]{1998A&A...337..832K} Kessel, O., Yorke, H.~W., Richling, S.\ 1998, \aap, 337, 832.% Photoevaporation of protostellar disks. III. The appearance of photoevaporating disks around young intermediate mass stars.

\bibitem[Keto 
\& Caselli(2008)]{2008ApJ...683..238K} Keto, E., \& Caselli, P.\ 2008, \apj, 683, 238 

\bibitem[Keto 
\& Caselli(2010)]{2010MNRAS.402.1625K} Keto, E., \& Caselli, P.\ 2010, \mnras, 402, 1625 

\bibitem[Kirk et al.(2007)]{2007ApJ...668.1042K} Kirk, H., Johnstone, D., 
\& Tafalla, M.\ 2007, \apj, 668, 1042.
%Dynamics of Dense Cores in the Perseus Molecular Cloud.

\bibitem[Krumholz et al.(2006)]{2006ApJ...653..361K} Krumholz, M.~R., 
Matzner, C.~D., \& McKee, C.~F.\ 2006, \apj, 653, 361.
%The Global Evolution of Giant Molecular Clouds. I. Model Formulation and Quasi-Equilibrium Behavior.

%\bibitem[Kudritzki, Yorke \& Frisch (1988)] {1988rmgm.book.....K} Kudritzki, R.~P., Yorke, H.~W., Frisch, H.\ 1988, 
%Radiation in moving gaseous media: Eighteenth Advanced Course of the Swiss Society of Astrophysics and Astronomy.

\bibitem[Kunz 
\& Mouschovias(2010)]{2010arXiv1003.2722K} Kunz, M.~W., \& Mouschovias, T.~C.\ 2010, arXiv:1003.2722 

\bibitem[Larson(1969)]{1969MNRAS.145..271L} Larson, R.~B.\ 1969, \mnras, 
145, 271.% Numerical calculations of the dynamics of collapsing proto-star. 

\bibitem[Lee et 
al.(1996)]{1996A&A...311..690L} Lee, H.-H., Herbst, E., Pineau des Forets, G., Roueff, E., \& Le Bourlot, J.\ 1996, \aap, 311, 690 

\bibitem[Lee et al.(2003)]{2003ApJ...583..789L} Lee, J.-E., Evans, N.~J., 
II, Shirley, Y.~L., \& Tatematsu, K.\ 2003, \apj, 583, 789. %Chemistry and Dynamics in Pre-protostellar Cores. 

\bibitem[Lee et al.(2004)]{2004ApJ...617..360L} Lee, J.-E., Bergin, E.~A., 
\& Evans, N.~J., II 2004, \apj, 617, 360. %Evolution of Chemistry and Molecular Line Profiles during Protostellar Collapse.

\bibitem[Li 
\& Shu(1996)]{1996ApJ...472..211L} Li, Z.-Y., \& Shu, F.~H.\ 1996, \apj, 472, 211 

\bibitem[Li(1999)]{1999ApJ...526..806L} Li, Z.-Y.\ 1999, \apj, 526, 806. 	
%A Spherical Model for Starless Cores of Magnetic Molecular Clouds and Dynamical Effects of Dust Grains.

\bibitem[Li et al.(2002)]{2002ApJ...569..792L} Li, Z.-Y., Shematovich, 
V.~I., Wiebe, D.~S., \& Shustov, B.~M.\ 2002, \apj, 569, 792. 
%A Coupled Dynamical and Chemical Model of Starless Cores of Magnetized Molecular Clouds. I. Formulation and Initial Results. 

%\bibitem[McCall et al.(2003)]{2003Natur.422..500M} McCall, B.~J., et al.\ 
%2003, \nat, 422, 500 

\bibitem[Mac Low 
\& Klessen(2004)]{2004RvMP...76..125M} Mac Low, M.-M., \& Klessen, R.~S.\ 2004, Reviews of Modern Physics, 76, 125.
% Control of star formation by supersonic turbulence. 

\bibitem[Mathis et al.(1977)]{mrn} Mathis, J.~S., Rumpl, 
W., \& Nordsieck, K.~H.\ 1977, \apj, 217, 425 

\bibitem[McDaniel \& Mason(1973)]{} McDaniel, E.~W., \& Mason, E.~A.\ 19737, in The
Mobility and Diffusion of Ions and Gases (New York: Wiley)

\bibitem[McKee 
\& Ostriker(2007)]{2007ARA&A..45..565M} McKee, C.~F., \& Ostriker, E.~C.\ 2007, \araa, 45, 565. %Theory of Star Formation.

\bibitem[Molek et al.(2009)]{2009IJMSp.285....1M} Molek, C.~D., Poterya, 
V., Adams, N.~G., 
\& McLain, J.~L.\ 2009, International Journal of Mass Spectrometry, 285, 1 

\bibitem[Morton, Mouschovias, \& Ciolek (1994)]{mmc94}
Morton, S. A., Mouschovias, T. Ch., \& Ciolek, G. E.\ 1994, ApJ, 421, 561

%\bibitem[Mouschovias(1991)]{1991ApJ...373..169M} Mouschovias, T.~C.\ 1991, 
%\apj, 373, 169 

\bibitem[Mouschovias \& Spitzer(1976)]{Mou1976} 
Mouschovias, T.~C., \& Spitzer, L., Jr.\ 1976, \apj, 210, 326 

\bibitem[Mouschovias 
\& Ciolek(1999)]{1999osps.conf..305M} Mouschovias, T.~C., \& Ciolek, G.~E.\ 1999, NATO ASIC Proc.~540: The Origin of Stars and Planetary Systems, 305. Magnetic Fields and Star Formation: A Theory Reaching Adulthood. 

\bibitem[Myers(1983)]{1983ApJ...270..105M} Myers, P.~C.\ 1983, \apj, 270, 
105. %Dense cores in dark clouds. III - Subsonic turbulence. 

\bibitem[Oberg et al.(2011)]{2011arXiv1107.5825O} Oberg, K.~I., Boogert, 
A.~C.~A., Pontoppidan, K.~M., van den Broek, S., van Dishoeck, E.~F., 
Bottinelli, S., Blake, G.~A., \& Evans, N.~J., II 2011, arXiv:1107.5825 


%\bibitem[Pavlyuchenkov et al.(2003)]{2003ARep...47..176P} Pavlyuchenkov, 
%Y.~N., Shustov, B.~M., Shematovich, V.~I., Wiebe, D.~S., 
%\& Li, Z.-Y.\ 2003, Astronomy Reports, 47, 176.% A Coupled, Dynamical and Chemical Model for the Prestellar Core L1544: Comparison of Modeled and Observed C18O, HCO+, and CS Emission Spectra.

\bibitem[Penston(1969)]{1969MNRAS.144..425P} Penston, M.~V.\ 1969, \mnras, 
144, 425. %Dynamics of self-gravitating gaseous spheres-III. Analytical results in the free-fall of isothermal cases.

%\bibitem[Preibisch et al.(1993)]{1993A&A...279..577P} Preibisch, T., Ossenkopf, V., Yorke, H.~W., Henning, T.\ 1993, \aap, 279, 577.%	The influence of ice-coated grains on protostellar spectra.

\bibitem[Rawlings et al.(1992)]{1992MNRAS.255..471R} Rawlings, J.~M.~C., 
Hartquist, T.~W., Menten, K.~M., \& Williams, D.~A.\ 1992, \mnras, 255, 471.% Direct diagnosis of infall in collapsing protostars. I - The theoretical identification of molecular species with broad velocity distributions.

%\bibitem[Rawlings 
%\& Yates(2001)]{2001MNRAS.326.1423R} Rawlings, J.~M.~C., \& Yates, J.~A.\ 2001, \mnras, 326, 1423 . %%Modelling line profiles in infalling cores.

%\bibitem[Redman et al.(2006)]{2006MNRAS.370L...1R} Redman, M.~P., Keto, E., 
%\& Rawlings, J.~M.~C.\ 2006, \mnras, 370, L1.% Oscillations in the stable starless core Barnard 68.

\bibitem[Scalo 
\& Elmegreen(2004)]{2004ARA&A..42..275S} Scalo, J., \& Elmegreen, B.~G.\ 2004, \araa, 42, 275. 	
%	Interstellar Turbulence II: Implications and Effects. 

\bibitem[Sch{\"o}ier et al.(2005)]{2005A&A...432..369S} Sch{\"o}ier, F.~L., van der Tak, F.~F.~S., van Dishoeck, E.~F., \& Black, J.~H.\ 2005, \aap, 432, 369 

\bibitem[Shematovich et al.(2003)]{2003ApJ...588..894S} Shematovich, V.~I., 
Wiebe, D.~S., Shustov, B.~M., \& Li, Z.-Y.\ 2003, \apj, 588, 894. %A Coupled Dynamical and Chemical Model of Starless Cores of Magnetized Molecular Clouds. II. Chemical Differentiation.

\bibitem[Shirley et al.(2000)]{2000ApJS..131..249S} Shirley, Y.~L., Evans, 
N.~J., II, Rawlings, J.~M.~C., \& Gregersen, E.~M.\ 2000, \apjs, 131, 249 

\bibitem[Schnee et al.(2010)]{2010ApJ...718..306S} Schnee, S., Enoch, M., 
Johnstone, D., Culverhouse, T., Leitch, E., Marrone, D.~P., 
\& Sargent, A.\ 2010, \apj, 718, 306 

\bibitem[Shu(1977)]{1977ApJ...214..488S} Shu, F.~H.\ 1977, \apj, 214, 488. 
Self-similar collapse of isothermal spheres and star formation. 

\bibitem[Smith et al.(2004)]{2004MNRAS.350..323S} Smith, I.~W.~M., Herbst, 
E., \& Chang, Q.\ 2004, \mnras, 350, 323 


\bibitem[Stantcheva et 
al.(2002)]{2002A&A...391.1069S} Stantcheva, T., Shematovich, V.~I., \& Herbst, E.\ 2002, \aap, 391, 1069 

\bibitem[Tafalla et al.(1998)]{1998ApJ...504..900T} Tafalla, M., Mardones, 
D., Myers, P.~C., Caselli, P., Bachiller, R., 
\& Benson, P.~J.\ 1998, \apj, 504, 900. %L1544: A Starless Dense Core with Extended Inward Motions. 

\bibitem[Tafalla et al.(2002)]{2002ApJ...569..815T} Tafalla, M., Myers, 
P.~C., Caselli, P., Walmsley, C.~M., \& Comito, C.\ 2002, \apj, 569, 815. %Systematic Molecular Differentiation in Starless Cores.

\bibitem[Tafalla et 
al.(2004)]{2004A&A...416..191T} Tafalla, M., Myers, P.~C., Caselli, P., \& Walmsley, C.~M.\ 2004, \aap, 416, 191.% On the internal structure of starless cores. I. Physical conditions and the distribution of CO, CS, N2H+, and NH3 in L1498 and L1517B. 

\bibitem[Tafalla 
\& Santiago(2004)]{2004A&A...414L..53T} Tafalla, M., \& Santiago, J.\ 2004, \aap, 414, L53. %L1521E: The first starless core with no molecular depletion. 

\bibitem[Tafalla et 
al.(2006)]{2006A&A...455..577T} Tafalla, M., Santiago-Garc{\'{\i}}a, J., Myers, P.~C., Caselli, P., Walmsley, C.~M., \& Crapsi, A.\ 2006, \aap, 455, 577 

\bibitem[Tan et al.(2006)]{2006ApJ...641L.121T} Tan, J.~C., Krumholz, 
M.~R., \& McKee, C.~F.\ 2006, \apjl, 641, L121.% Equilibrium Star Cluster Formation. 

\bibitem[Tassis 
\& Mouschovias(2005)]{2005ApJ...618..769T} Tassis, K., \& Mouschovias, T.~C.\ 2005, \apj, 618, 769 

\bibitem[Tassis 
\& Mouschovias(2007)]{2007ApJ...660..388T} Tassis, K., \& Mouschovias, T.~C.\ 2007, \apj, 660, 388. 
%Protostar Formation in Magnetic Molecular Clouds beyond Ion Detachment. II. Typical Axisymmetric Solution. 

\bibitem[Tassis 
\& Yorke(2011)]{ty10} Tassis, K., \& Yorke, H.~W.\ 2011, \apjl, 735, L32 

\bibitem[Terzieva 
\& Herbst(1998)]{1998ApJ...501..207T} Terzieva, R., \& Herbst, E.\ 1998, \apj, 501, 207 
%The Sensitivity of Gas-Phase Chemical Models of Interstellar Clouds to C and O Elemental Abundances and to a New Formation Mechanism for Ammonia.


%\bibitem[van Leer (1979)]{vL79}
%van Leer, B.\ 1979, J. Comput. Phys., 32, 101

\bibitem[van Weeren et 
al.(2009)]{2009A&A...497..773V} van Weeren, R.~J., Brinch, C., \& Hogerheijde, M.~R.\ 2009, \aap, 497, 773 


\bibitem[Vasyunin et al.(2009)]{2009ApJ...691.1459V} Vasyunin, A.~I., 
Semenov, D.~A., Wiebe, D.~S., \& Henning, T.\ 2009, \apj, 691, 1459.
%A Unified Monte Carlo Treatment of Gas-Grain Chemistry for Large Reaction Networks. I. Testing Validity of Rate Equations in Molecular Clouds. 

%\bibitem[Velusamy et al.(2008)]{2008ApJ...688L..87V} Velusamy, T., Peng, 
%R., Li, D., Goldsmith, P.~F., \& Langer, W.~D.\ 2008, \apjl, 688, L87.
%Dichotomy in the Dynamical Status of Massive Cores in Orion.

\bibitem[Wakelam et 
al.(2006)]{2006A&A...451..551W} Wakelam, V., Herbst, E., \& Selsis, F.\ 2006, \aap, 451, 551 

\bibitem[Wakelam et 
al.(2010)]{2010A&A...517A..21W} Wakelam, V., Herbst, E., Le Bourlot, J., Hersant, F., Selsis, F., \& Guilloteau, S.\ 2010, \aap, 517, A21 

\bibitem[Walmsley(1991)]{1991IAUS..147..161W} Walmsley, M.\ 1991, 
Fragmentation of Molecular Clouds and Star Formation, 147, 161 

\bibitem[Ward-Thompson et al.(1994)]{1994MNRAS.268..276W} Ward-Thompson, 
D., Scott, P.~F., Hills, R.~E., \& Andre, P.\ 1994, \mnras, 268, 276 

\bibitem[Ward-Thompson et al.(1999)]{1999MNRAS.305..143W} Ward-Thompson, 
D., Motte, F., \& Andre, P.\ 1999, \mnras, 305, 143. 
%The initial conditions of isolated star formation - III. Millimetre continuum mapping of pre-stellar cores.

\bibitem[Whittet(2003)]{2003dge..conf.....W} Whittet, D.~C.~B.\ 2003, Dust 
in the galactic environment, 2nd ed.~ by D.C.B.~Whittet.~Bristol: Institute 
of Physics (IOP) Publishing, 2003 Series in Astronomy and Astrophysics.

\bibitem[Whittet et al.(2009)]{2009ApJ...695...94W} Whittet, D.~C.~B., 
Cook, A.~M., Chiar, J.~E., Pendleton, Y.~J., Shenoy, S.~S., 
\& Gerakines, P.~A.\ 2009, \apj, 695, 94 

\bibitem[Willacy(2007)]{2007ApJ...660..441W} Willacy, K.\ 2007, \apj, 660, 
441. %The Chemistry of Multiply Deuterated Molecules in Protoplanetary Disks. I. The Outer Disk. 

\bibitem[Woodall et 
al.(2007)]{2007A&A...466.1197W} Woodall, J., Ag{\'u}ndez, M., Markwick-Kemper, A.~J., \& Millar, T.~J.\ 2007, \aap, 466, 1197. %The UMIST database for astrochemistry 2006. 

\bibitem[Xie et al.(1995)]{1995ApJ...440..674X} Xie, T., Allen, M., 
\& Langer, W.~D.\ 1995, \apj, 440, 674 

%\bibitem[Yorke (1988)]{ 1988rmgm.conf..195Y}	 Yorke, H.~W.\ 1988, in Radiation in moving gaseous media: eighteenth Advanced Course of the Swiss Society of Astrophysics and Astronomy, Kudritzki, Yorke, Frisch, p. 195, Radiation in Diffuse Matter.

\bibitem[Yorke \& Kr\"ugel (1977)] {1977A&A....54..183Y} Yorke, H.~W. \& Kr\"ugel, E.\ 1977, \aap, 54, 183. %The Dynamic Evolution of a Massive Protostar Envelope.

%\bibitem[Yorke, Tenorio-Tagle, Bodenheimer (1984)]{1984A&A...138..325Y} Yorke, H.~W., Tenorio-Tagle, G. \& Bodenheier, P.\ 1984 \aap, 138, 325. Line formation in H II regions during the champagne phase.

%\bibitem[Zinnecker \& Yorke (2007)]{2007ARA&A..45..481Z} Zinnecker, H. \& Yorke, H.~W.\ 2007,  \araa, 45, 481. %Toward Understanding Massive Star Formation.


\end{thebibliography}
\end{document}